\documentclass[twocolumn,notitlepage,prb,superscriptaddress,longbibliography]{revtex4-2}
\usepackage{amsfonts}
\usepackage{textcomp}
\usepackage{times}
\usepackage{graphicx}
\usepackage{float}
\usepackage{latexsym,amsmath,amssymb,bm,euscript}
\usepackage{color}
\usepackage{subfigure}
\usepackage[colorlinks=true,linkcolor=blue,citecolor=blue,urlcolor=blue]{hyperref}
\usepackage{hyperref}
\usepackage{soul}
\usepackage[normalem]{ulem}
\usepackage{mathrsfs}
\usepackage{amsmath}

\begin{document}
\preprint{APS2014}
\title[]{Ground-state phase diagram of the extended two-leg $t$-$J$ ladder}
\author{Xin Lu}
\affiliation{School of Physics, Beihang University, Beijing 100191, China}
\author{Dai-Wei Qu}
\affiliation{Kavli Institute for Theoretical Sciences, University of Chinese Academy of Sciences, Beijing 100190, China}
\affiliation{CAS Center for Excellence in Topological Quantum Computation, University of Chinese Academy of Sciences, Beijing 100049, China}
\author{Yang Qi}
\email{qiyang@fudan.edu.cn}
\affiliation{State Key Laboratory of Surface Physics and Department of Physics, Fudan University, Shanghai 200433, China}
\affiliation{Collaborative Innovation Center of Advanced Microstructures, Nanjing 210093, China}
\author{Wei Li}
\email{w.li@itp.ac.cn}
\affiliation{CAS Key Laboratory of Theoretical Physics, Institute of Theoretical Physics, 
Chinese Academy of Sciences, Beijing 100190, China}
\affiliation{CAS Center for Excellence in Topological Quantum Computation, University of Chinese Academy of Sciences, Beijing 100049, China}
\affiliation{Peng Huanwu Collaborative Center for Research and Education, Beihang University, Beijing 100191, China}
\author{Shou-Shu Gong}
\email{shoushu.gong@buaa.edu.cn}
\affiliation{School of Physics, Beihang University, Beijing 100191, China}

\date{\today}

\begin{abstract}
Inspired by the observation of a robust $d$-wave superconducting phase driven by tuning the next-nearest-neighbor (NNN) electron hopping in recent density matrix renormalization group (DMRG) studies of six- and eight-leg $t$-$J$ model, we systematically study the phase diagram of the two-leg $t$-$J$ ladder with the NNN couplings (including NNN hopping and spin interaction) in a large region of doping level, by means of the DMRG calculations. 
Upon doping from half filling, we identify the Luther-Emery liquid (LEL) phase, which can be distinguished as the pairing-dominant and charge density-dominant regime by comparing the Luttinger parameter $K_{\rho}$. 
With the growing NNN couplings, pairing correlations are enhanced and correspondingly $K_{\rho}$ increases, driving the system from the density-dominant to the pairing-dominant regime. 
In the Tomonaga-Luttinger liquid (TLL) phase in the larger doping region, we identify two TLL regimes with different features of charge density correlation. 
At the quarter filling ($1/2$ doping level), we find that the strong dimer orders of bond energy in the open system actually decay algebraically and thus do not indicate a spontaneous translational symmetry breaking.
Our results show that in the LEL phase of the two-leg ladder, the NNN couplings seem to play the similar role as that on the wider $t$-$J$ cylinder, 
and studies on this more accessible system can be helpful towards understanding the emergence of the remarkable $d$-wave superconducting phase on the wider system.
\end{abstract}

\maketitle

\section{\label{sec:ITR} INTRODUCTION}

Understanding the emergence of the unconventional superconductivity (SC) is one of the major challenges in condensed matter physics.
Since the unconventional SC is usually realized by doping the parent antiferromagnetic Mott insulators, the doped Hubbard and $t$-$J$ models are usually taken as the canonical models for studying the SC in strongly correlated systems~\cite{Science.367.6475,Science.235.4793,PhysRevB.35.8865,PhysRevLett.61.2376,PhysRevB.39.11413,PhysRevLett.98.067006,RevModPhys.78.17,PhysRevLett.59.2095}.
Although the microscopic pairing mechanism in doped Mott insulators remains elusive, it is believed that slightly doping Mott insulators can lead to an unconventional SC. 

As a first step towards the two-dimensional (2D) doped Mott insulators, the two-leg Hubbard~\cite{PhysRevB.46.3159,PCS.270.281,PhysRevB.102.115136,PhysRevB.48.15838,PhysRevB.53.12133,PhysRevB.56.6569} and $t$-$J$ ladders~\cite{PhysRevB.53.11721,PhysRevB.102.104512,PhysRevB.76.195105,PhysRevB.72.014523,PhysRevB.52.6796,PhysRevB.95.245105,PhysRevB.65.165122,PhysRevLett.101.217001,PhysRevB.60.13418} have been studied extensively and have been well understood by combining density matrix renormalization group (DMRG) and bosonization calculations~\cite{PhysRevB.65.165122,PhysRevB.53.R2959,PhysRevB.56.7167,PhysRevB.51.3709}.
A typical state in such systems is the Tomonaga-Luttinger liquid (TLL)~\cite{QPIOD}, which has gapless spin and charge sectors as well as consequent algebraic correlation functions.
More interestingly, for small doping ratio the spin sector may be gapped but the charge sector remains gapless, leading to the Luther-Emery liquid (LEL)~\cite{PhysRevLett.33.589,PhysRevB.71.045113,PhysRevB.68.052504} state with algebraic pairing and charge density correlation functions.
This LEL state has the $d$-wave pairing symmetry and is considered as a quasi-one-dimensional (1D) SC state~\cite{PhysRevB.64.100506}, which therefore has been taken as a basic model to construct and test effective theories to understand the emergent SC in the electronic systems with repulsive interaction~\cite{PhysRevB.89.144501,arXiv:2203.05480,arXiv:2202.01320}.
Since the ground state of the two-leg ladder at half filling has a finite spin gap and short-range spin correlation~\cite{PhysRevB.60.13418,PhysRevB.58.1794,PhysRevLett.73.886}, the resonating valence bond (RVB) theory provides a straightforward picture to understand the pairing~\cite{Science.235.4793}, in which the doped holes favor to bind into pairs to minimize energy~\cite{PhysRevB.57.11666,PhysRevB.53.251}.
The phase string theory~\cite{PhysRevLett.77.5102,PhysRevB.55.3894,IJMPB2007}, which was built upon a singular phase string effect induced by the motion of holes in a doped Mott insulator, has also been numerically tested to explain both the pairing of two holes and the emergent LEL phase in two-leg $t$-$J$ models~\cite{PhysRevB.102.104512,PhysRevB.98.245138}. 
With growing system circumference, a natural question is that whether the LEL state can develop to the $d$-wave SC in two dimensions. 

In recent years, by extensive numerical simulations on the wider square-lattice Hubbard (large $U$) and $t$-$J$ models, it has been found that near the optimal doping the $d$-wave SC pairing correlations decay exponentially and a stripe order emerges~\cite{PhysRevLett.80.1272,Science.358.1155,PhysRevLett.91.136403,PhysRevB.60.R753,PhysRevB.97.045138,PhysRevB.71.075108,PhysRevB.95.125125,PhysRevB.100.195141,PhysRevX.10.031016}.
Very recently, DMRG calculations found that in the slight doping regime of the $t$-$J$ model, the next-nearest-neighbor (NNN) hopping $t_2$ ($t_2/t_1 > 0$ and $t_1$ is the nearest-neighbor (NN) hopping) can suppress the stripe order and lead to a robust $d$-wave SC phase~\cite{PhysRevLett.127.097003, PhysRevLett.127.097002, PNAS.118.44,PhysRevB.106.174507}. While the positive $t_2 / t_1$ may increase the mobility of pairs~\cite{PhysRevB.58.9492,PhysRevB.60.R753,PhysRevB.79.220504,PhysRevB.64.180513,PhysRevB.63.014414}, theoretical understanding of the enhanced SC with tuning hopping is not very clear.
Considering it is more controllable to develop theories on two-leg ladder, investigating the SC with doping and tuning $t_2/t_1$ on two-leg ladder will be helpful towards understanding of the role of the NNN hopping, which however has not been studied systematically to the best of our knowledge.

Specifically, previous studies on the two-leg $t$-$J$ ladder have been focused on the systems without NNN couplings.
For the isotropic $t$-$J$ ladder at low doping level, the system is in the LEL phase at small $J_1/t_1$, and with growing $J_1 / t_1$ a phase separation occurs~\cite{PhysRevB.65.165122,PhysRevLett.101.217001}. 
For small $J_1 / t_1$ such as $J_1 / t_1 = 1/3$, the system is in charge gapped states at the commensurate doping levels $\delta =1/4$ ($3/8$ filling) and $\delta = 1/2$ (quarter filling)~\cite{PhysRevB.65.165122,PhysRevLett.101.217001}.
While the charge gapped state at $\delta =1/4$ is a charge density wave (CDW) state breaking translational symmetry~\cite{PhysRevB.63.195106}, it was suggested to be a bond order wave at $\delta =1/2$~\cite{PhysRevB.76.195105,PhysRevLett.101.217001}. 
With further growing doping level, the system enters the TLL phase~\cite{PhysRevB.58.3425,PhysRevB.52.6796,PhysRevB.53.11721}, but the properties in the TLL phase have not been carefully studied.

\begin{figure}
   \includegraphics[width=0.48\textwidth,angle=0]{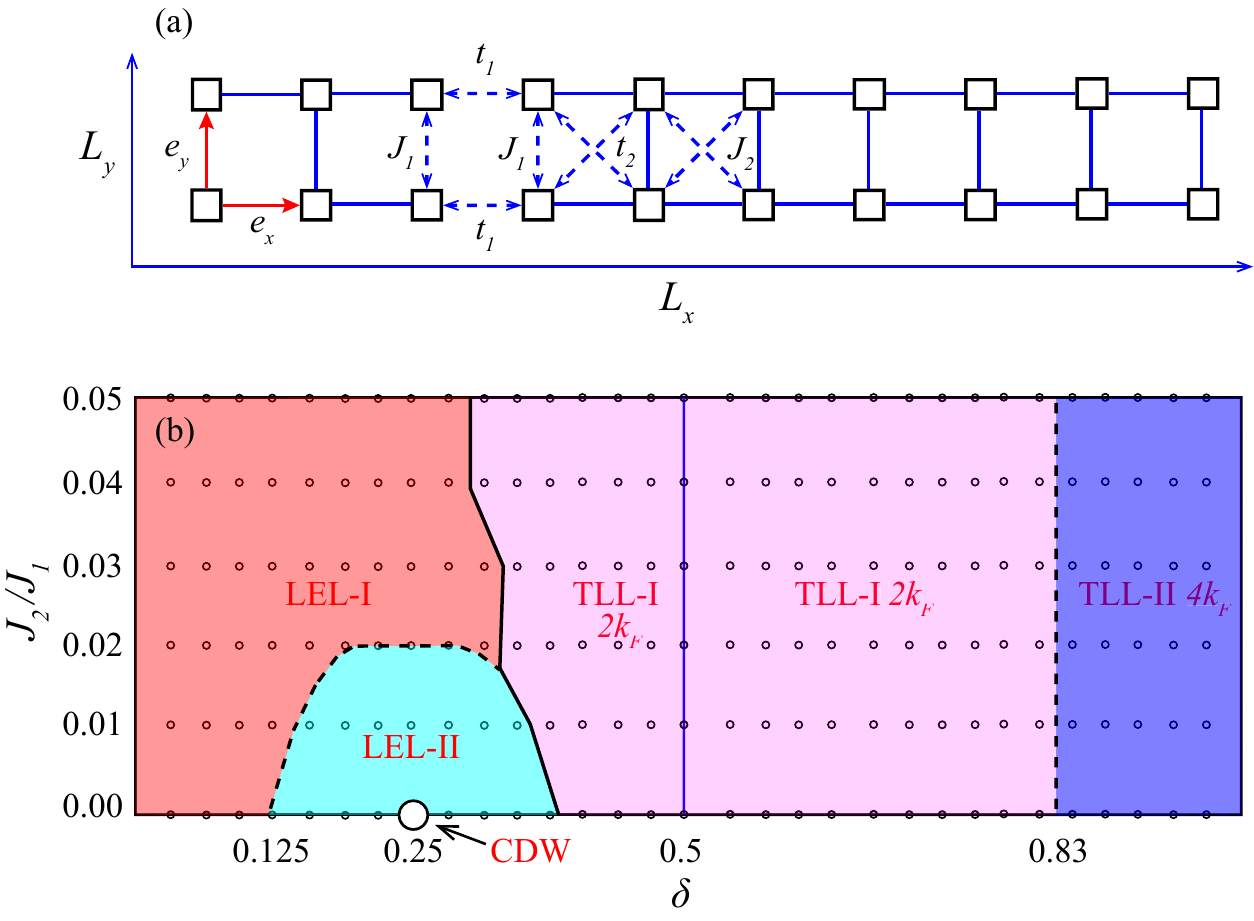}
   \caption{\label{Geo-Pha}
   Schematic figure of the two-leg $t$-$J$ ladder and the phase diagram.
   (a) The two-leg $t$-$J$ ladder with the nearest-neighbor and next-nearest-neighbor hopping $t_1, t_2$ and spin interactions $J_1, J_2$. The open boundary conditions are imposed along the ${\mathit{\mathbf{e}}}_x$ direction. $L_x$ and $L_y = 2$ are the numbers of sites in the two directions. 
   (b) Quantum phase diagram of the model obtained on the $L_x=128$ and $L_y=2$ lattice in the range of $0\le J_2 /J_1 \le 0\ldotp 05$ and $0<\delta <0.9$. $(t_2 / t_1)^2 = J_2 / J_1$ is fixed with tuning parameter and $t_{1}/J_{1}=3.0$ is chosen in this work. The small black points denote the parameters we have calculated using DMRG. We identify a Luther-Emery liquid (LEL) and a Tomonaga-Luttinger liquid (TLL) phase. The LEL phase is denoted as two regimes, the LEL-I regime with dominant pairing correlation and the LEL-II regime with dominant charge density correlation. The TLL phase is also distinguished as two regimes, which show quite different features of the structure factor of charge density correlation. The fully gapped charge density wave (CDW) at $\delta=1/4$ is represented by a black hollow circle and the C0S1 state at $\delta=1/2$ is represented by a solid blue line. The phase boundary between the LEL and TLL is denoted as a solid line, and the different regimes in the same phase are separated by dashed line.}
\end{figure}

In this paper, we carefully study the two-leg $t$-$J$ ladder with the NNN couplings (including NNN hopping and spin interaction) $t_2, J_2$ (see Fig.~\ref{Geo-Pha}(a)) from small to large doping regime, by means of DMRG calculations.
We study the $t$-$J$ ladder with the doping level up to $\delta = 0.9$ in the presence of the NNN couplings and our results are summarized in Fig.~\ref{Geo-Pha}(b).
In the LEL phase upon doping, the pairing correlations are also enhanced by the NNN couplings, consistent with the previous findings on the wider systems~\cite{PhysRevResearch.2.033073}.
Since the increased doping level suppresses pairing correlations, we identify the pairing-dominant (LEL-I) and charge density-dominant (LEL-II) regime by comparing the Luttinger parameter $K_{\rho}$.
The CDW state at $\delta = 1/4$ is driven to the LEL by very small NNN couplings, which is consistent with the tiny charge gap in the absence of the NNN couplings~\cite{PhysRevB.65.165122}.
In the TLL phase, we find that sub-leading peaks of the density correlation structure factor change from the momentum $(2 k_F, 0)$ in the so-dubbed TLL-I regime to $(4 k_F, 0)$ in the TLL-II regime, where $k_F$ is the Fermi momentum of free electrons with the same filling.
At the commensurate $1/2$ doping, we find that the bond order wave proposed in previous study~\cite{PhysRevB.76.195105} is actually not long-ranged
and the ground state does not break the translational symmetry.
Our results show that the enhanced SC by tuning the NNN hopping ($t_2/t_1 > 0$) is a rather general conclusion in the square-lattice $t$-$J$ model, which suggests that developing theory for two-leg ladder can be helpful for understanding the emergent $d$-wave SC on the wider systems.

The paper is organized as follow.
In Sec.~\ref{sec:MAM}, we introduce the model Hamiltonian and the details of DMRG calculations. In Sec.~\ref{sec:LEL}, we study the LEL phase by tuning NNN couplings and doping levels. We also analyze the transition from a fully gaped CDW state to the LEL at $1/4$ doping. In Sec.~\ref{sec:TLL}, we discuss the properties of the TLL phase, and also point out the absent long-ranged bond order at $1/2$ doping. The last section Sec.~\ref{sec:CON} is devoted to the summary and discussion.

\section{\label{sec:MAM} Model and method}

The Hamiltonian of the $t$-$J$ model is defined as:
\begin{equation}\label{H-tJ}
     H = -\sum_{\{ij\},\sigma}t_{ij}(\hat{c}^{\dagger}_{i,\sigma} \hat{c}_{j,\sigma} + h.c.)  + \sum_{\{ij\}} J_{ij} (\hat{\bf S}_i \cdot \hat{\bf S}_j - \frac{1}{4} \hat{n}_i \hat{n}_j),
\end{equation}
where ${\hat{c} }_{i\sigma }^{\dagger }$ and ${\hat{c} }_{i\sigma }$ are respectively the creation and annihilation operators for the electrons with spin $\sigma $ $(\sigma = \pm 1/2)$ on the site $i=(x_i,y_i)$. 
${\hat{\bf S} }_i$ is the spin-$1/2$ operator and ${\hat{n} }_i =\sum_{\sigma } {\hat{c} }_{i\sigma }^{\dagger } {\hat{c} }_{i\sigma }$ is the electron number operator. The Hilbert space for each site is constrained to no-double occupancy. We consider the interactions between both NN and NNN sites, and denote the corresponding hopping (spin interaction) as $t_1$ ($J_1$) and $t_2$ ($J_2$).
The length and width of the ladder are $L_x$ and $L_y$ ($\equiv 2$), giving the total site number $N = L_x \times L_y$.
We study hole-doped case with doping ratio $\delta$ defined as $\delta = N_h / N$, where $N_h$ is the number of doped holes.
Below we choose $J_1 = 1.0$ as the energy unit and set $t_1 /J_1 = 3$.
We study the ground state of the system with doping level up to $\delta = 0.9$ by tuning $J_2 / J_1$ up to $0.05$, with $t_2 / t_1$ correspondingly up to $0.22$ under the constraint $(t_2 / t_1)^2 = J_2 / J_1$ to make a connection to the corresponding Hubbard model~\cite{PhysRevB.95.075124}.

We solve the ground state of the system by DMRG calculations~\cite{PhysRevLett.69.2863,PhysRevB.48.10345}.
We use the open boundary conditions in the $x$ direction and choose $L_x = 128$ for most calculations (and up to $L_x = 256$). 
We keep up to $1200$ $\rm SU(2)$ multiplets (equivalent to about $3600$ U(1) states) in DMRG calculations to ensure the truncation error smaller than ${10}^{-8}$ (see Ref.~\cite{PhysRevLett.127.097003} for the details of the SU(2) DMRG algorithm of the $t$-$J$ model). 
For longer systems with $L_x$ up to $256$, we keep bond dimensions up to $3000$ to ensure a full convergence.

To characterize many-electron states, we calculate four types of correlation functions and analyze their structure factors. The spin-spin correlation function is defined as
\begin{equation}
   F\left(r\right)=\frac{1}{L_y }\sum_{y=1}^{L_y } \left\langle {\hat{\bf S}}_{\left(x_0 ,y\right)} \cdot {\hat{\bf S}}_{\left(x_0 +r,y\right)} \right\rangle,
\end{equation}
where ${\hat{\bf S}}_{(x ,y)}$ is the spin operator at the site $(x,y)$, and $(x_0,y)$ is the reference site.  
The single-particle Green's function reads
\begin{equation}
   G \left(r\right)=\frac{1}{L_y }\sum_{\sigma,y=1}^{L_y } \left\langle \hat{c}_{\left(x_0 ,y\right),\sigma }^{\dagger } \hat{c}_{\left(x_0 +r,y\right),\sigma } \right\rangle,
\end{equation}
and the charge-density correlation function is expressed as
\begin{equation}
D\left(r\right)=\frac{1}{L_y }\sum_{y=1}^{L_y } \left\langle {\hat{n} }_{(x_0 ,y)} {\hat{n} }_{(x_0 +r,y)} \right\rangle -\left\langle {\hat{n} }_{(x_0 ,y)} \right\rangle \left\langle {\hat{n} }_{(x_0 +r,y)} \right\rangle.
\end{equation}
For characterizing superconductivity we calculate the singlet pairing correlation function
\begin{equation}\label{SC-LE}
   \Phi_{\alpha \beta } \left(r\right)=\left\langle \hat{\Delta}_{\alpha }^{\dagger } \left(x_0 ,y\right) \hat{\Delta}_{\beta } \left(x_0 +r,y\right)\right\rangle,
\end{equation}
where ${\hat{\Delta}}_{\alpha }^{\dagger }(x,y)$=$({\hat{c} }_{(x,y)\uparrow }^{\dagger } {\hat{c} }_{(x,y)+{\mathit{\mathbf{e}}}_{\alpha }\downarrow }^{\dagger } -{\hat{c} }_{(x,y)\downarrow }^{\dagger } {\hat{c} }_{(x,y)+{\mathit{\mathbf{e}}}_{\alpha }\uparrow }^{\dagger })/\sqrt{2}$
is the spin-singlet pair-field creation operator, and ${\mathit{\mathbf{e}}}_\alpha $ ($\alpha = x,y$) denotes the unit length along $x$ or $y$ direction. 
In most parameter region of our study, the different types of pairing correlation functions have the similar power exponent except in the TLL-I region with $\delta < 0.5$, where the power exponent of $\Phi_{x x}(r)$ is slightly smaller than others (see Appendix~\ref{d-wave} for the comparison of the different pairing correlations).
Therefore, we mainly demonstrate $\Phi_{y y} \left(r\right)$ in the main text, which characterizes the pairing correlations between the vertical bonds.
To diminish the boundary effect, we study the correlation functions by choosing the reference site with $x_0 \sim L_x / 4$, and the corresponding structure factors are obtained by taking the Fourier transformation
\begin{equation}
   Q\left(\vec{k}\right)=\frac{1}{N}\sum_{i,j} Q\left(\vec{r}_i, \vec{r}_j\right)e^{i\vec{k} \cdot \left({\vec{r} }_i -{\vec{r} }_j \right)},
\end{equation}
where $Q\left(\vec{r}_i, \vec{r}_j \right)$ is the correlation function in real space.

In the following parts, we will fit the correlation functions to estimate their power exponents.
We will fit the data points with relatively large magnitudes, which gives small error range and thus we will not show the error bar in the figure.

\section{\label{sec:LEL} Luther-Emery Liquid and Charge Density Wave}

Firstly, we discuss the LEL and CDW at the lower doping side.
The LEL state has a finite spin gap but vanishing charge gap, denoted as the C1S0 state, which means there is a gapless mode in the charge (C) sector and no gapless mode in the spin (S) sector~\cite{PhysRevB.53.12133}. 
Therefore, the LEL has an exponentially decaying spin correlation function and a finite central charge $c = 1$ corresponding to the gapless charge mode. 
In the charge-2e sector, both the density and pairing correlation functions exhibit the algebraic decaying behaviours. 
Based on the bosonization theory for two-leg ladder systems~\cite{QPIOD,PhysRevB.63.195106,PhysRevB.12.3908,PhysRevB.102.104512}, pairing and density correlations in the LEL phase follow
\begin{equation}\label{LE-Phi}
  \Phi \left(r\right)=\frac{A_0 }{r^{1/(2K_{\rho })} }+A_1 \frac{\mathrm{c}\mathrm{o}\mathrm{s}\left(2k_F r\right)}{r^{2K_{\rho } + 1/(2K_{\rho })} },
\end{equation}  
\begin{equation}\label{LE-D}
   D\left(r\right)=-\frac{K_{\rho } }{{\left(\pi r\right)}^2 }+B_1 \frac{\mathrm{cos}\left(2k_F r\right)}{r^{2K_{\rho } } },
\end{equation}
where the parameters $A_0$, $A_1$ and $B_1$ depend on model details. $K_{\rho}$ is the Luttinger parameter that controls the decaying exponents of correlation functions and $k_F$ is the Fermi momentum related to the electron filling number.
Usually, the pairing correlation $\Phi \left(r\right)$ in Eq.~\eqref{LE-Phi} is dominated by the first term $r^{-K_{\mathrm{sc}} } \sim r^{-1/(2K_{\rho }) }$ and the density correlation  $D\left(r\right)$ in Eq.~\eqref{LE-D} is dominated by the second term $r^{-K_{\mathrm{c}} } \sim r^{-2K_{\rho } }$, giving the characteristic feature of the LEL $K_{\mathrm{sc}} \cdot K_{\mathrm{c}} \simeq 1$.

We start with the $1/4$ doping case. Previous numerical calculations and bosonization analysis have identified a fully gapped CDW state with a four-fold degeneracy~\cite{PhysRevLett.101.217001,PhysRevB.63.195106,PhysRevB.76.195105} and very small spin and charge gaps~\cite{PhysRevB.65.165122,PhysRevB.95.245105}.
This fully gapped state can be also understood in the bosonization theory. 
At this commensurate doping level the Umklapp and backscattering processes are relevant, thus the sine-Gordon type Hamiltonian density will be appended with a set of cosine-type terms which describe the scattering processes between different Fermi points in the bonding and anti-bonding bands~\cite{fradkin_2013,PhysRevB.102.115136,PhysRevB.68.052504,PhysRevB.84.054517,ScienceBulletin.63.753},
\begin{equation}\label{BKT}
   \mathcal{H} = \frac{v_{\mu } }{2}\left\lbrack K_{\mu } \Pi_{\mu }^2 +\frac{1}{K_{\mu } }{\left(\partial_x \phi_{\mu } \right)}^2 \right\rbrack +V_{\mu } \mathrm{c}\mathrm{o}\mathrm{s}\left(\beta \sqrt{8}\pi \phi_{\mu } \right),
\end{equation}
where $\mu =\rho ,\sigma$ represent respectively the charge and spin degrees of freedom, with the Einstein notation assumed. 
$v_{\mu }$ are the renormalized Fermi velocities, $\phi_{\mu }$ are the bosonic fields describing spin and charge excitations, $\Pi_{\mu }$ are the conjugate field operators for $\phi_{\mu }$, and $V_{\mu }$ stand for the interactions. 
In the cosine-type terms, $\beta$ is a commensurate factor related to the specific filling case, and $2\beta$ represents the ground-state degeneracy~\cite{Motttransition,PhysRevB.65.153110}.
For $\delta = 1/4$, we have $\beta = 2$ accounting for the four-fold degeneracy.
Next, we study the LEL and CDW states in presence of the NNN couplings.

\subsection{Correlation functions in the Luther-Emery liquid phase}

\begin{figure*}
   \includegraphics[width=1.0\textwidth,angle=0]{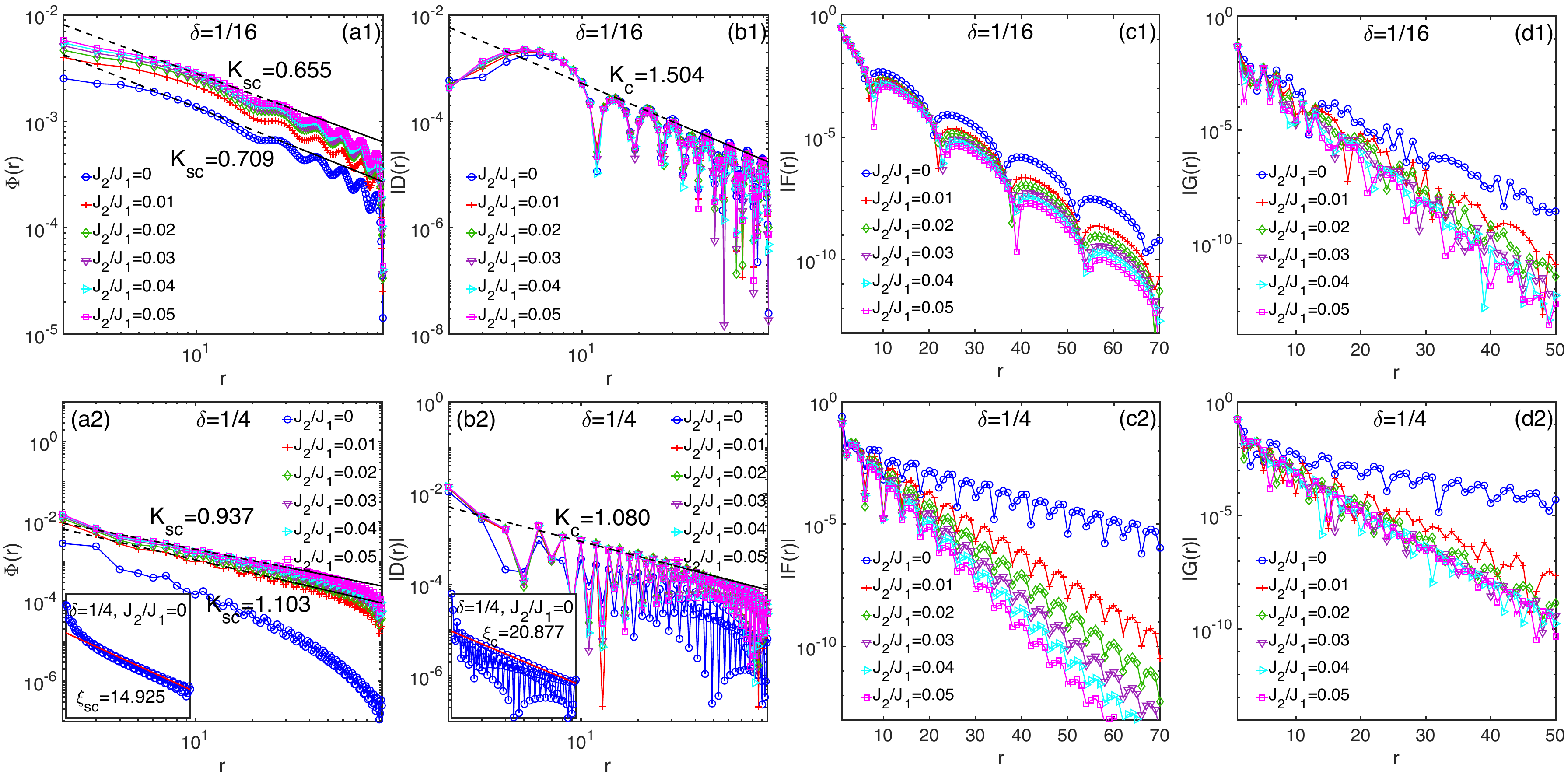}
   \caption{\label{Correfunc-LEL}
   Correlation functions in the LEL phase. Two commensurate doping ratios $\delta=1/16$ ($N_{h}=16$) and $\delta=1/4$ ($N_{h}=64$) are chosen as representatives. (a1-d1) are pairing correlation, density correlation, spin correlation and single-particle Green's function for 
   $\delta=1/16$ with different $J_2/J_1$. The power exponents $K_{\mathrm{sc}}$ and $K_{\mathrm{c}}$ are extracted by fitting the strongest points of the pairing and density correlation functions. (a1) and (b1) are double-logarithmic scale. (c1) and (d1) are semi-logarithmic scale. (a2-d2) are the similar figures for $\delta=1/4$. The special point at $J_{2}/J_{1}=0$ and $\delta=1/4$ is a fully gapped charge density wave state. The insets in (a2) and (b2) are the pairing and density correlation functions for $J_{2}/J_{1}=0$, $\delta=1/4$ with semi-logarithmic scale. The fittings give the large correlation lengths $\xi_{\mathrm{sc}}=14.925$ and $\xi_{\mathrm{c}}=20.877$, respectively.}
\end{figure*}

We first study correlation functions both in real space and momentum space (see Appendix~\ref{Stru-fac-LEL}) to characterize the LEL. 
We show the results for two doping ratios $\delta = 1/16$ and $1/4$ as representatives in Fig.~\ref{Correfunc-LEL}. For the given doping ratio, pairing correlations are enhanced by increasing $J_2 / J_1$ and $K_{\rm sc}$ also decreases slightly. 
We have confirmed that the power exponents of the pairing and density correlations satisfy $K_{\mathrm{sc}} \cdot K_{\mathrm{c}} \simeq 1$.
On the other hand, if we fix $J_{2} / J_{1}$ and increase doping level, $K_{\mathrm{sc}}$ increases and $K_{\mathrm{c}}$ decreases, showing that doping will suppress pairing correlation but enhance density correlation. 
The quantitative discussion on the dependence of these power exponents on $J_2 / J_1$ and $\delta$ will be presented in subsection~\ref{Lutt-para}.
Furthermore, we show spin correlations and single-particle Green's functions in Figs.~\ref{Correfunc-LEL}(c1)-\ref{Correfunc-LEL}(c2) and Figs.~\ref{Correfunc-LEL}(d1)-\ref{Correfunc-LEL}(d2), respectively.
Both correlations decay exponentially and faster with the increase of $J_2 / J_1$, which support the gapped nature of the LEL and suggest the spin and single-particle excitation gaps increase as $J_2/J_1$ increase.

In the LEL phase, correlation function show a modulation in the decay (vs. $r$) whose period is doping ratio dependent.
For the pairing and absolute value of spin correlation the periods is $\lambda = 1/\delta$, and for density correlation it is $\lambda = 1 / 2\delta$ (after taking the absolute value).
The spin correlations still show the $\pi$-phase shift~\cite{PhysRevLett.80.1272,PhysRevB.79.220504,PhysRevB.60.R753,PhysRevLett.81.3227}.
In Fig.~\ref{Correfunc-LEL}, one finds that the NNN couplings do not change the oscillation periods of correlation functions in the LEL phase.

\subsection{Charge density profile and central charge in the Luther-Emery liquid phase}\label{central-charge-1}

\begin{figure*}
   \includegraphics[width=0.9\textwidth,angle=0]{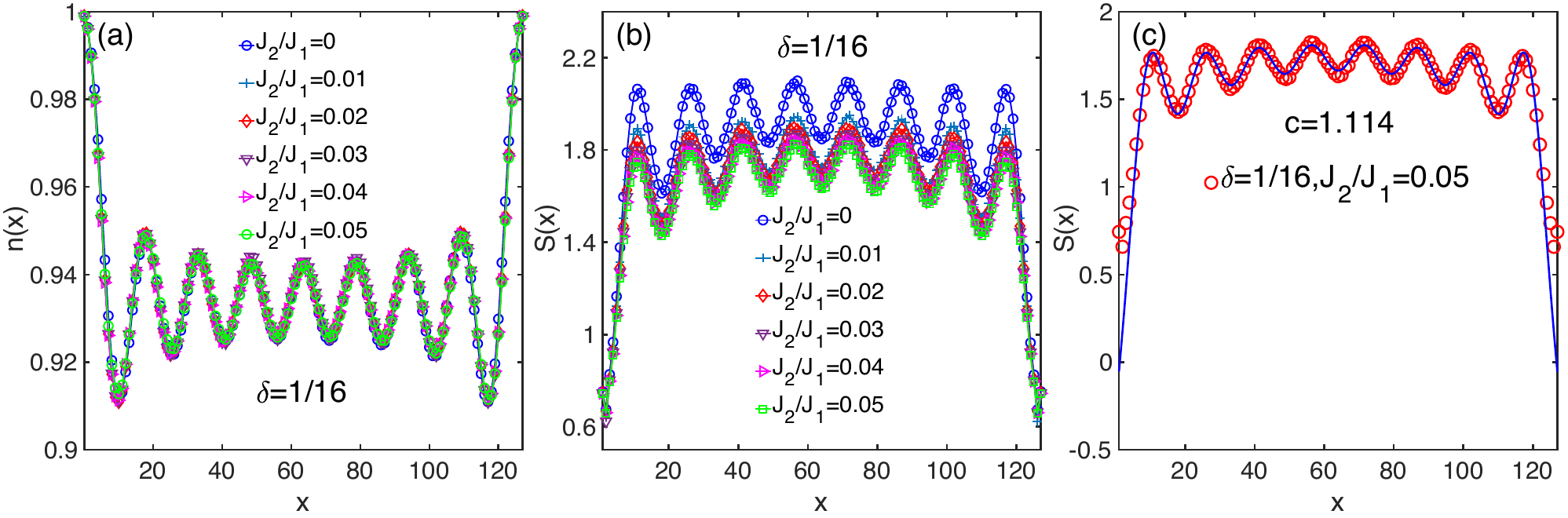}
   \caption{\label{CDW-entropy}
   Charge density profile and central charge in the LEL phase. The data at $\delta=1/16$ are demonstrated as an example. (a) Charge density profile $n(x)$ for each rung with the position $x$. (b) Entanglement entropy $S(x)$ versus the subsystem length $x$. (c) The fitting of the central charge from the entanglement entropy by Eq.~\eqref{entropy}. The obtained fitting parameters are $c=1.114$, $q=0.207$, $A=-2.492$ and $B=1.084$.}
\end{figure*}

For the LEL state on the two-leg ladder with OBC, charge density has the Friedel oscillation induced by the open boundaries~\cite{PhysRevB.65.165122}.
We denote the total charge density in each rung as $n(x)$, where $x$ labels the position of the rung.
The charge density distribution is described by 
\begin{equation}
n\left(x\right)=n_0 +A_{\mathrm{cdw}} \mathrm{cos}\left(\frac{4\pi}{\lambda} x+\phi \right),
\label{eq:cdw_fit} 
\end{equation}
where $A_{\mathrm{cdw}}=A_{0}[x^{-K_c /2} +{\left(L_x +1-x\right)}^{-K_c /2}]$~\cite{PhysRevLett.127.097003,science.aal5304}, in which $K_c$ characterizes the algebraic decay of the density amplitude from the boundaries to the bulk, with $L_x$ the length of the ladder.
We examine the charge density in the LEL phase and show the results in Fig.~\ref{CDW-entropy}(a), 
where the charge density changes only slightly in the studied region of $J_2 / J_1$, for a fixed doing ratio of $\delta=1/16$, and the period of charge density profile satisfies the LEL feature $\lambda = 1/\delta$.
We use Eq.~\eqref{eq:cdw_fit} to fit the density profiles and extract the power exponents $K_c$ (c.f. Appendix~\ref{Kc-CDW}), which agree with those obtained from fitting the algebraic decay of density correlations in Fig.~\ref{Correfunc-LEL}(b1).

To further examine the nature of the LEL, we calculate entanglement entropy and extract the central charge.
The entanglement entropies $S\left(x\right)=-\mathrm{Tr}\left\lbrack \hat{\rho}_x \mathrm{ln}\hat{\rho}_x \right\rbrack$ are shown in Fig.~\ref{CDW-entropy}(b), where $\hat{\rho}_x$ is the reduced density matrix of the subsystem with rung number $x$.
The entropy data also have an oscillation with a wavelength $\lambda = 1 / \delta$, the same as the charge density oscillations.
For a 1D critical system described by the conformal field theory, it has been established~\cite{CFT1,CFT2} that on an open boundary system with length $L_x$, the entanglement entropy follows
\begin{align}\label{entropy}
  \begin{split}
        S\left(x\right)&=\frac{c}{6}\mathrm{ln}\left\lbrack \frac{4\left(L_x +1\right)}{\pi }\mathrm{sin}\frac{\pi \left(2x+1\right)}{2\left(L_x +1\right)}\left|\mathrm{sin}\;\frac{2\pi}{\lambda}\right|\right\rbrack \\
     & +\frac{A\;\mathrm{sin}\left\lbrack \frac{2\pi}{\lambda}\left(2x+1\right)\right\rbrack }{\frac{4\left(L_x +1\right)}{\pi }\mathrm{sin}\frac{\pi \left(2x+1\right)}{2\left(L_x +1\right)}\left|\mathrm{sin}\;\frac{2\pi}{\lambda}\right|}+B,
  \end{split}
\end{align}
where $A, B$ are model-dependent parameters, $2\pi / \lambda$ approaches the Fermi momentum $k_F$ in the thermodynamic limit, and the second term describes the contribution from the higher-order oscillation~\cite{PDW}. 
Taking the entropy data of $J_{2}/J_{1}=0.05$ in Fig.~\ref{CDW-entropy}(c) as an example, we use Eq.~(\ref{entropy}) to fit the center charge.
To avoid the boundary effect in the fitting, a few data points near the boundaries are excluded in the fitting. 
The entropy is fitted quite well by a central charge $c = 1.114$ consistent with a gapless charge mode.
The entropies for the other cases in Fig.~\ref{CDW-entropy}(b) can be similarly fitted that also find the central charge $c\approx 1$.

\subsection{$K_{{\rm sc}}$-dominant and $K_{{\rm c}}$-dominant regimes in the Luther-Emery liquid phase}\label{Lutt-para}

\begin{figure}
   \includegraphics[width=0.48\textwidth,angle=0]{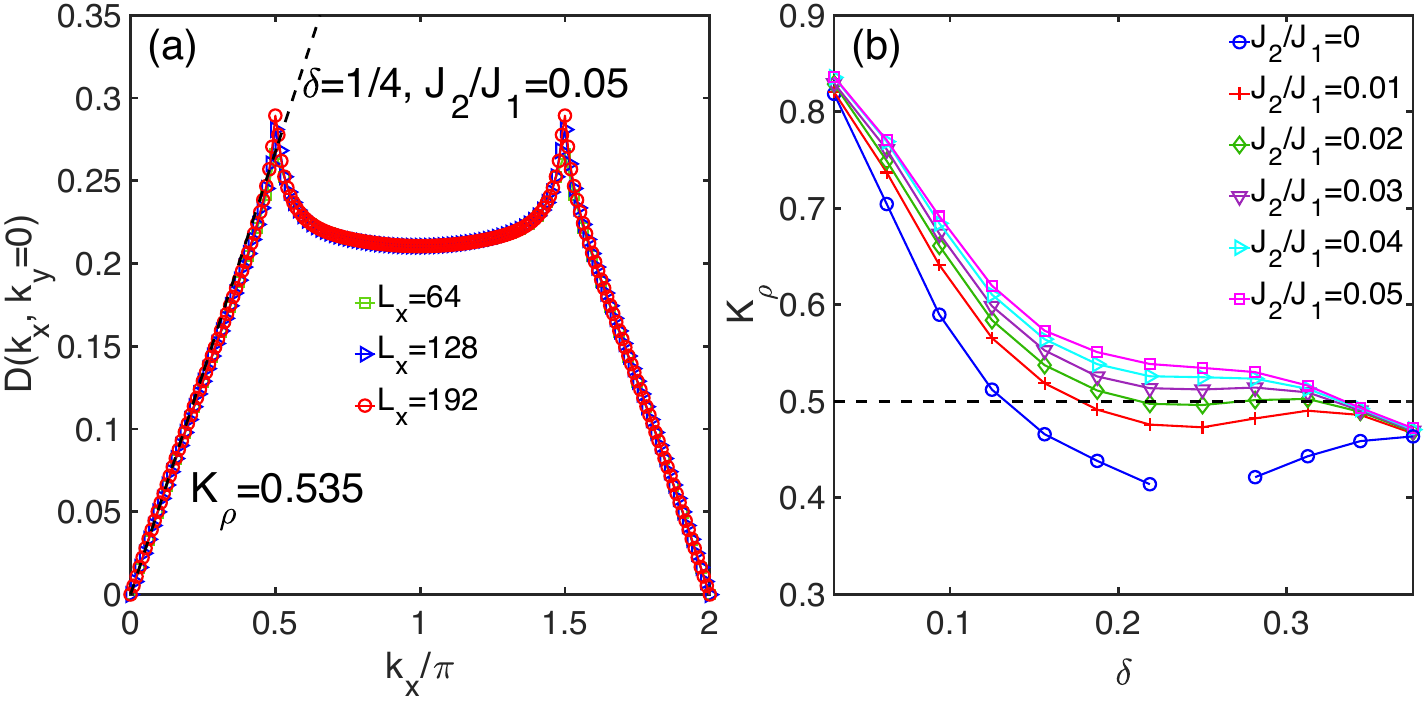}
   \caption{\label{Two-type-LEL}
   Determination of the Luttinger parameter $K_{\rho}$ in the LEL phase.
   (a) The structure factors of density correlation function $D(k_x, k_y = 0)$ at $\delta=1/4$ and $J_{2}/J_{1}=0.05$ with different system sizes $L_x = 64, 128, 192$. Fitting the slope near the momentum point $(0,0)$ by Eq.~\eqref{krho} gives $K_{\rho} = 0.535$. (b) The obtained Luttinger parameter $K_{\rho}$ in the LEL phase, at different doping levels and $J_2/J_1$. The system at $J_{2}/J_{1}=0$, $\delta=1/4$ is a gapped CDW state without a well defined $K_{\rho}$. The dotted line denotes $K_{\rho}=0.5$.}
\end{figure}

By comparing the correlation functions in Fig.~\ref{Correfunc-LEL}, one can find that the $t_2$, $J_2$ can enhance the pairing correlation but the increased doping level suppresses it.
Since $K_{{\rm sc}} \cdot K_{{\rm c}} \simeq 1$ in the LEL, the different $K_{{\rm sc}}$-dominant and $K_{{\rm c}}$-dominant behaviors may exist with tuning parameters,  distinguished by the Luttinger parameter $K_{\rm \rho}$ as $K_{{\rm sc}} = 1/(2K_{\rho })$ and $K_{{\rm c}} = 2K_{\rm \rho}$. 
$K_{\rm \rho} > 1/2$ indicates $K_{{\rm sc}}$ dominant, otherwise it is $K_{{\rm c}}$ dominant.
From Eq.~\eqref{LE-D}, one can obtain the Luttinger parameter from density structure factor as 
\begin{equation}\label{krho}
   K_{\rho } =\pi \underset{\mathbf{k}\to 0^+ }{\mathrm{lim}} \frac{D\left(\mathbf{k}\right)}{\mathbf{k}}=\pi \underset{\mathbf{k}\to 0^+ }{\mathrm{lim}} \frac{1}{\mathbf{k}}\frac{1}{N}\sum_r e^{-i\mathbf{k}\cdot \mathbf{r}} D\left(\mathbf{r}\right),
\end{equation}  
which indicates $K_{\rho}$ is proportional to the slope of $D\left(\mathbf{k}\right)$ at $\mathbf{k} = (0,0)$ and thus can be extracted~\cite{PhysRevB.83.205113,PhysRevB.59.4665,EPL.70.492,EPL.87.27001}. In Fig.~\ref{Two-type-LEL}(a), we show the calculated $K_{\rho}$ at $\delta = 1/4$, $J_{2}/J_{1}=0.05$.
The density structure factors for different lattice sizes lay on top of each other, from which the Luttinger parameter $K_{\rho }$ can be extracted.

With this we obtain $K_{\rho}$ for various parameter points in the LEL phase, as shown in Fig.~\ref{Two-type-LEL}(b) with doping level starting from $\delta = 1/32$.
By comparing $K_{\rho}$ with $0.5$~\cite{PhysRevB.65.165122,PhysRevB.53.R2959}, we determine the boundary between the $K_{{\rm sc}}$-dominant and $K_{{\rm c}}$-dominant regimes in Fig.~\ref{Geo-Pha}(b).
Notice that the condition $K_{\rho} < 0.5$ is also satisfied in the TLL phase, which can be distinguished from the $K_{{\rm c}}$-dominant LEL by their different central charges. 
In Fig.~\ref{Two-type-LEL}(b), $K_{\rho}$ at $\delta = 1/4, J_{2}/J_{1}=0$ is not shown since this parameter point corresponds to a CDW state without a well-defined Luttinger parameter~\cite{PhysRevB.65.165122,QPIOD}.

\subsection{Phase transition from the charge density wave to the Luther-Emery liquid at $1/4$ doping}

\begin{figure*}
   \includegraphics[width=0.9\textwidth,angle=0]{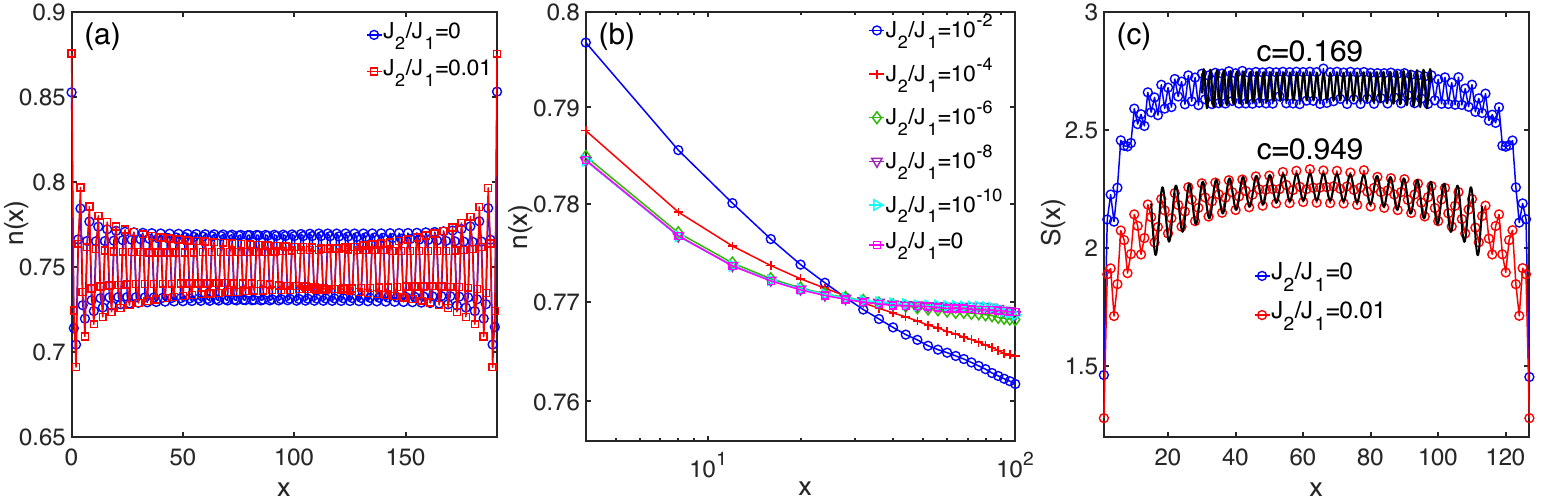}
   \caption{\label{Destroyed-CDW}
   Charge density profiles and entanglement entropies at $1/4$ doping. (a) Charge density profile $n(x)$ for $J_2/J_1 = 0$ and $0.01$. (b) Double-logarithmic plots of the upper density values of $n(x)$ versus $x$, for different $J_{2}/J_{1}$. (c) Fittings of the central charge from entanglement entropy for $J_2/J_1 = 0$ and $0.01$ on the $L_x = 128$ ladder. For $J_2/J_1 = 0$, it is a small value close to zero. For $J_2/J_1 = 0.01$, it is close to $1$.}
\end{figure*}

In the previous studies of the conventional $t$-$J$ ladder without $t_2$, $J_2$, a fully gapped CDW state with small gaps has been found at the commensurate $\delta = 1/4$ for the case of $t_{1}/J_{1}=3$~\cite{PhysRevB.65.165122,PhysRevLett.101.217001,PhysRevB.63.195106,PhysRevB.76.195105}, which can be witnessed by the exponential decay correlation functions with large correlation lengths, as shown in the insets of Figs.~\ref{Correfunc-LEL}(a2) and \ref{Correfunc-LEL}(b2). 
Below, we study the phase transition from the CDW to the LEL with growing $J_2/J_1$.

We first study the charge density profile.
In the open-boundary system, while there exists truly long-range charge order in the CDW phase, it show algebraic quasi-long range order in the LEL phase~\cite{PhysRevB.65.165122}.
One example for system length $L_x = 192$ is shown in Fig.~\ref{Destroyed-CDW}(a), 
from which we find that while the charge density oscillation for $J_2/J_1 = 0$ has a robust amplitude in the bulk, the amplitude for $J_2/J_1 = 0.01$ decays from the boundary to the bulk.
To show the difference more clearly, we plot the upper density values versus their positions in the double-logarithmic scale, as shown in Fig.~\ref{Destroyed-CDW}(b). 
It is clear that while the upper density values are flat in the truly long-range CDW phase, they decay algebraically in the LEL phase.

Since the CDW and LEL state have different numbers of gapless modes, we can also locate the transition by computing the central charge $c$ from the entanglement entropy. 
One example is shown in Fig.~\ref{Destroyed-CDW}(c) for $J_2 / J_1 = 0$ and $0.01$.
We fit the entropy using Eq.~\eqref{entropy} and skip the data points near both ends to avoid the boundary effect.
The fittings give $c \approx 0$ for $J_2 / J_1 = 0$ and $c \approx 1.0$ for $J_2 / J_1 = 0.01$, which agree with the CDW and LEL, respectively. 
One may understand this transition from the bosonization theory.
While both $V_{\rho}$ and $V_{\sigma }$ are relevant in this fully gapped CDW phase~\cite{PhysRevB.65.165122,PhysRevB.71.045113}, the Umklapp processes ($V_{\rho }$) are no longer relevant but the backscattering ($V_{\sigma }$) processes still contribute in the LEL phase~\cite{backward,PhysRevB.68.052504}. 
This transition occurs between the charge-gapless and gapped phase, which is described as a Kosterlitz-Thouless (KT) transition by the sine-Gordon model~\cite{BSCS,JPAMG.13.1980}.
In Appendix~\ref{central-charge and entropy}, we demonstrate more results of the fitted central charge and entanglement entropy to show this phase transition.

\section{\label{sec:TLL} Tomonaga-Luttinger Liquid and C0S1 state at the quarter filling}

With further increase of doping ratio, the bosonization theory predicts that the backscattering ($V_{\sigma }$) processes will be suppressed gradually, which can lead to a transition from the LEL to the TLL with spin gap closing~\cite{QPIOD,fradkin_2013}.
Therefore, correlation functions decay algebraically in the TLL phase and can be obtained from the bosonization theory~\cite{fradkin_2013,BSCS,QPIOD,PhysRevB.83.205113}. 
For the two-leg ladder with $t_{\perp } \le t_{\parallel }$, the key difference of correlation functions from those of the TLL in one-dimensional chain is the Luttinger parameter, which was suggested to change from $K_{\rho }$ to $2K_{\rho }$~\cite{PhysRevB.53.11721} and has been confirmed~\cite{PhysRevB.53.R2959,PhysRevB.49.16078,PhysRevB.51.16456}.

Following the previous results of the TLL~\cite{PhysRevB.53.11721,PhysRevB.83.205113}, we first list the bosonization predictions of different correlation functions here. The singlet pairing correlation function defined in Eq.~\eqref{SC-LE} has two important algebraic modes, following the behavior
\begin{equation}\label{TL-Phi}
   \Phi \left(r\right)=\frac{C_0 }{r^{1+ 1/(2K_{\rho }) } }+C_1 \frac{\mathrm{cos}\left(2k_F r\right)}{r^{2K_{\rho } + 1/(2K_{\rho }) } },
\end{equation}
where $C_0$ and $C_1$ depend on model details.
For density correlation function, there are three important terms given as
\begin{equation}\label{TL-D}
   D\left(r\right)=-\frac{K_{\rho } }{{\left(\pi r\right)}^2 }+D_1 \frac{\mathrm{cos}\left(2k_F r\right)}{r^{1+2K_{\rho } } }{\mathrm{ln}}^{-\frac{3}{2}} \left(r\right)+D_2 \frac{\mathrm{cos}\left(4k_F r\right)}{r^{8K_{\rho } } }.
\end{equation}
Meanwhile, the spin correlation function follows the behavior
\begin{equation}\label{TL-F}
   F\left(r\right)=\frac{E_0}{{\left(\pi r\right)}^2 }+E_1 \frac{\mathrm{cos}\left(2k_F r\right)}{r^{1+2K_{\rho } } }{\mathrm{ln}}^{\frac{1}{2}} \left(r\right),
\end{equation}
and the single-particle Green's function always shows a algebraic decay form $G\left(r\right)\thicksim r^{-1-\alpha }$, where
\begin{equation}\label{TL-G}
   \alpha =2\sum_{\mu } \frac{1}{8}\left(K_{\mu } +\frac{1}{K_{\mu } }-2\right),\;\;\;\mu =\rho ,\sigma.
\end{equation}
In presence of spin rotational SU(2) symmetry, $K_{\sigma}=1$~\cite{SU(2)-K}.
At the commensurate quarter filling, the NN $t$-$J$ model is a C0S1 state. 
We will also reexamine this state in this section.

\subsection{TLL-I regime}

\begin{figure*}
   \includegraphics[width=1.0\textwidth,angle=0]{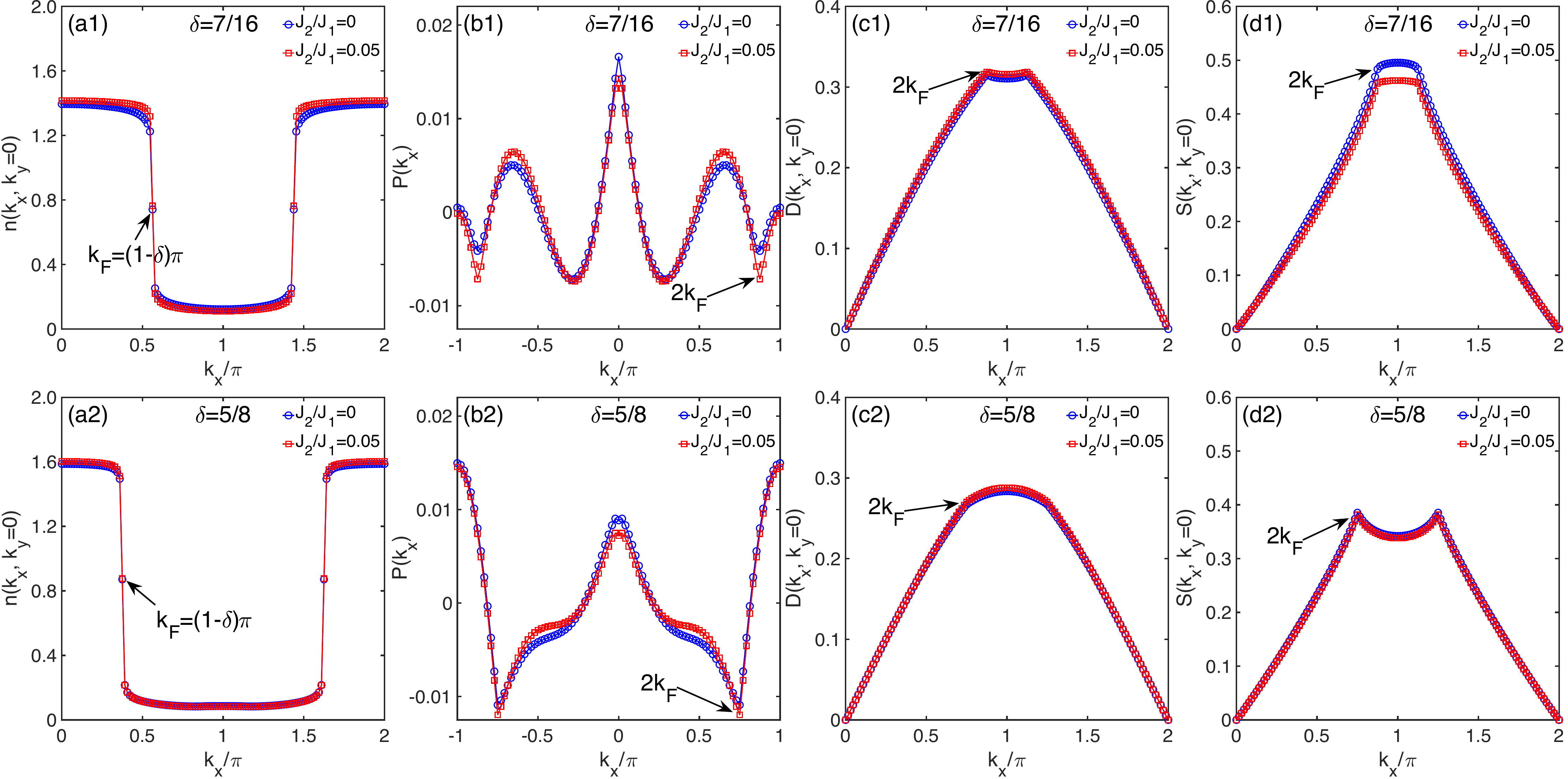}
   \caption{\label{Structure-factors-TLL}
   Structure factors in the TLL-I regime.
   Two representative doping ratios $\delta=7/16$ and $\delta=5/8$ with $J_2/J_1 = 0, 0.05$ are selected to demonstrate the results for $\delta<0.5$ and $\delta>0.5$, respectively.
   (a1)-(d1) are the electron density in momentum space, pairing correlation structure factor, density correlation structure factor, and spin structure factor at $\delta = 7/16$. (a2)-(d2) are the same figures for $\delta=5/8$. Some symbolic peaks and kinks are marked in the figures, which are related to the Fermi momentum $k_F$.}
\end{figure*}

By checking correlation functions and fitting the central charge, we identify the TLL-I regime for both $\delta < 0.5$ and $\delta > 0.5$ (see Fig.~\ref{Geo-Pha}(b)). 
In Fig.~\ref{Structure-factors-TLL}, we show the results in the momentum space at $\delta = 7/16$ and $\delta = 5/8$, below and above the quarter filling. 
Since the correlation functions change slightly with growing $J_2 / J_1$ in our studied region, we only discuss the results for $J_{2}/J_{1} = 0$ and $0.05$ as representatives, 
which are also shown in real space in Appendix~\ref{correlation-TLL}.

We first determine the Fermi momentum $k_F$. 
Since electrons in one-dimensional systems are strongly correlated, the electron density in momentum space has no discontinuity at $k_F$ but shows a power-law behavior away from $k_F$~\cite{PhysRevB.104.035149}
\begin{equation}
   n\left(k\right)=n\left(k_F \right)+A \mathrm{sign}\left(k\pm k_F \right) {\left|k\pm k_F \right|}^{\alpha },
\end{equation}
where the parameter $\alpha$ can be obtained from Eq.~\eqref{TL-G}.
In two-leg ladders, the Fermi momentum strictly meets the constraint of the Luttinger sum rule $k_F = k_{F}^b + k_{F}^a = (1-\delta)\pi$, where $b$ and $a$ denote the bonding and anti-bonding bands~\cite{PhysRevB.76.195105,PCS.270.281,PhysRevB.95.155116}. 
We show $n(k_x, k_y = 0)$ in Figs.~\ref{Structure-factors-TLL}(a1) and \ref{Structure-factors-TLL}(a2), in which $n(k)$ changes rapidly at $k_F$.
The values of $k_F$ seem to be independent of $J_2 / J_1$, in consistent with the relation $k_F = (1-\delta) \pi$.

In Figs.~\ref{Structure-factors-TLL}(b1) and \ref{Structure-factors-TLL}(b2), we show the structure factor $P(k_x)$ of the pairing correlation $\Phi_{yy}(r)$.
Since $\Phi_{yy}(r)$ is along the $x$ direction, the Fourier transform $P(k_x)$ has only one momentum variable $k_x$.
For $\delta = 7/16$ in the $\delta < 0.5$ side, $P(k_x)$ has two sharp peaks at $k_x = 0$ and $2k_F$, indicating that both the uniform and $2k_F$ mode in Eq.~\eqref{TL-Phi} contribute to the pairing correlation function.
For $\delta = 5/8$ in the $\delta > 0.5$ side, the peak of $P(k_x)$ at $k_x = 0$ is weakened and slightly splits, which we confirm for other parameter points in this region.
Nonetheless, the contribution of the $2k_F$ mode is still significant, and we have also confirmed that $K_{\mathrm{sc}}$ is close to the power exponent of the $2 k_F$ mode, which is $2K_{\rho }+ 1/(2K_{\rho })$.

In Figs.~\ref{Structure-factors-TLL}(c1) and \ref{Structure-factors-TLL}(c2), we show the density structure factors $D(k_x, k_y = 0)$, which possess one of the main features in this TLL-I regime: besides the uniform component in Eq.~\eqref{TL-D}, the $2 k_F$ mode has a significant contribution to the density correlation.
Notice that the singular behaviors of the structure factors near $k_x = 2 k_F$ can be consistent with the logarithmic correction ${\mathrm{ln}}^{-3/2} \left(r\right)$ in Eq.~\eqref{TL-D}.
However, in the spin structure factors shown in Figs.~\ref{Structure-factors-TLL}(d1) and \ref{Structure-factors-TLL}(d2), the singular behaviors near $k_x = 2 k_F$ are much weaker than the expectation from the ${\mathrm{ln}}^{1/2} \left(r\right)$ correction, which indicates that the logarithmic correction in the second term of Eq.~\eqref{TL-F} may not properly describe the spin correlation in this TLL-I regime.

\subsection{The C0S1 state at the quarter filling}

\begin{figure}
   \includegraphics[width=0.48\textwidth,angle=0]{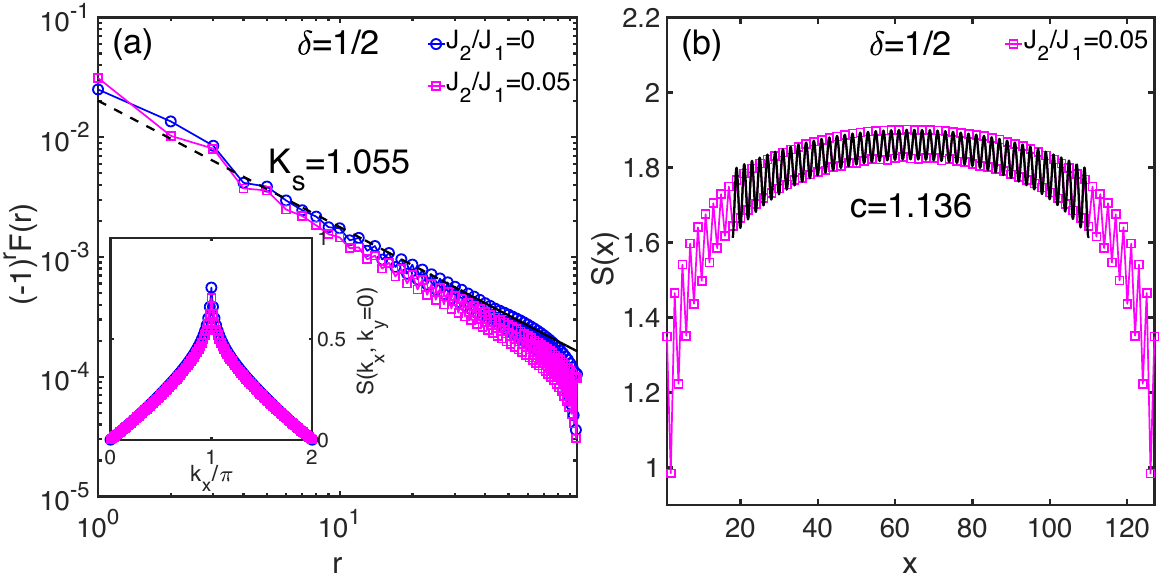}
   \caption{\label{R-SDW}
   Spin correlation function and central charge in the C0S1 state at $1/2$ doping. (a) Spin correlation functions in the double-logarithmic scale for $J_2/J_1 = 0, 0.05$. The power exponent $K_{\mathrm{s}}=1.055$ is extracted by fitting the data of $J_{2}/J_{1}=0$. The inset shows the corresponding spin structure factors for the $k_y=0$ component. (b) Fitting the central charge for $J_{2}/J_{1}=0.05$ by Eq.~\eqref{entropy}. The obtained fitting parameters are $c=1.136$, $q=1.572$, $A=-6.519$ and $B=0.895$. In the fitting, some data points near the boundaries have been omitted to avoid boundary effects. For other $J_{2}/J_{1}$ at $\delta = 1/2$, the central charge is also close to $1$.}
\end{figure}

At the commensurate doping $\delta = 1/2$ with $J_{2}/J_{1}=0$, previous numerical and analytic studies have identified a C0S1 state~\cite{PhysRevB.53.12133,PhysRevB.76.195105,PhysRevLett.101.217001}.
Under the OBC, the charge density is uniform but the NN horizontal bond energies show strong dimer oscillations, which was interpreted as a signal of a bond order wave~\cite{PhysRevB.76.195105}.

\begin{figure*}
   \includegraphics[width=0.9\textwidth,angle=0]{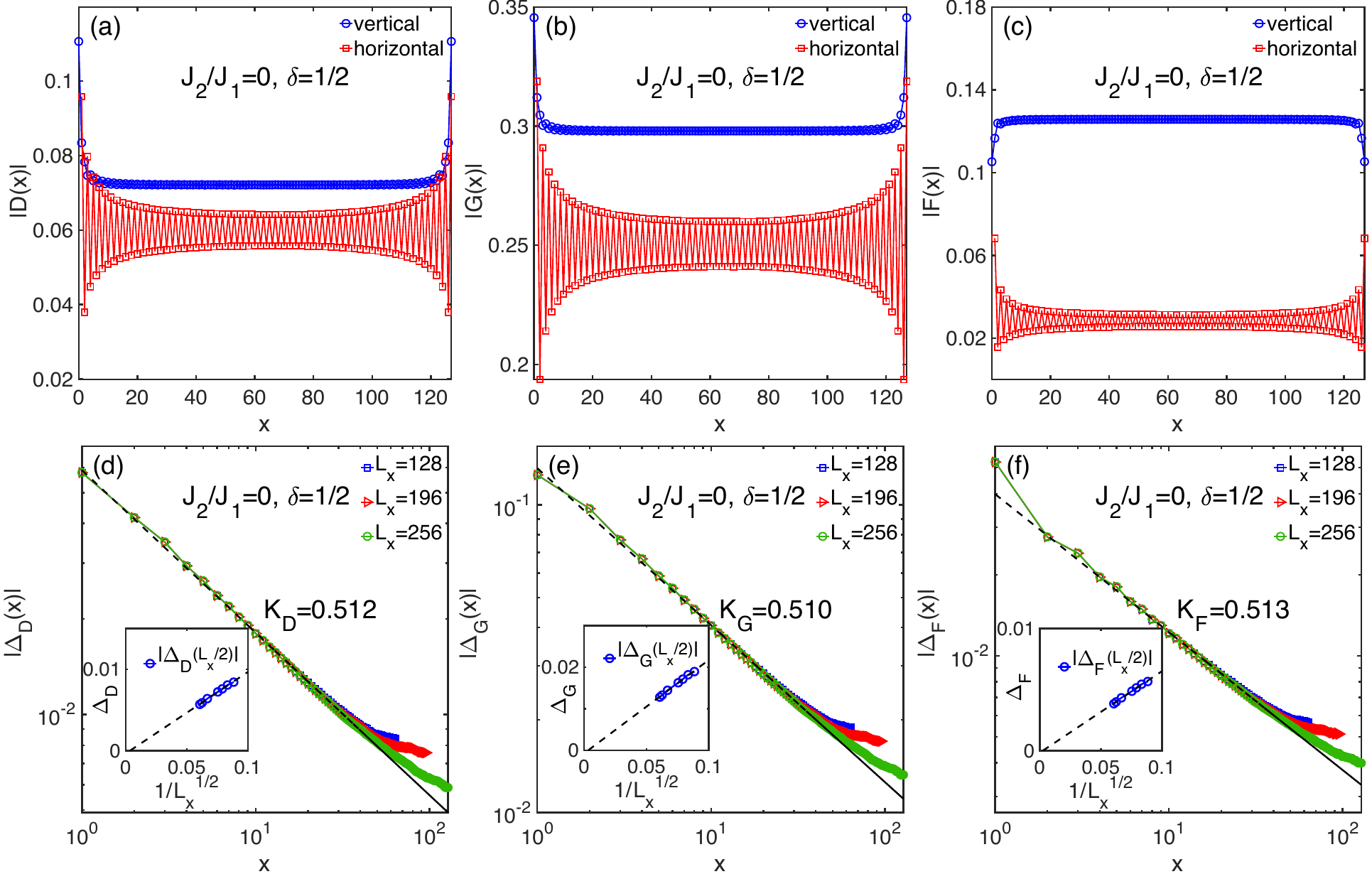}
   \caption{\label{BOW}
   Vanished dimer order of the C0S1 state at the $1/2$ doping level.
   (a)-(c) are the nearest-neighbor vertical and horizontal density correlation, single-particle correlation, and spin correlation, respectively. (d-f) Double-logarithmic plots of the corresponding dimer order parameters on different system sizes. The insets show the dimer order parameters in the middle of the ladder ($x=L_x /2$) versus the system length $1/\sqrt{L_x}$.}
\end{figure*}

With growing $J_2/J_1$ at $\delta = 1/2$, we first investigate the spin correlation as shown in Fig.~\ref{R-SDW}(a).
We find that spin correlations follow the algebraic decay for different $J_{2}/J_{1}$, and the power exponents are all close to $1$, consistent with gapless spin excitations.
The structure factors shown in the inset have a peak at $\mathbf{k} = (\pi, 0)$ and characterize the antiferromagnetic spin correlation along the chain direction.
Furthermore, we check the central charge.
in Fig.~\ref{R-SDW}(b), where the fitting of the entanglement entropy at $J_{2}/J_{1}=0.05$, leads to $c \approx 1$ and indicates that up to $J_{2}/J_{1}=0.05$, the system is still in the C0S1 state.

In the previous DMRG study of the $t$-$J$ ladder at the quarter filling, the NN horizontal bond energies have strong dimer oscillations, implying a possible bond order wave that spontaneously breaks translational symmetry and with two-fold degenerate ground states~\cite{PhysRevB.76.195105}.
Here, we reexamine this possible order by simulating a much longer system with length $L_x$ up to 256.
In Figs.~\ref{BOW}(a)-\ref{BOW}(c), we show $\langle \hat{n}_i \hat{n}_j\rangle$, $\sum_{\sigma} \langle \hat{c}^{\dagger}_{i,\sigma} \hat{c}_{j,\sigma} + h.c. \rangle$ and $\langle {\hat{\bf S}}_i \cdot {\hat{\bf S}}_j \rangle$ for all the NN horizontal and vertical bonds at $J_{2}/J_{1}=0$. 
Similar to the previous results~\cite{PhysRevB.76.195105}, the vertical bond energies are uniform in the bulk, but the horizontal bond energies show strong oscillations.
To determine whether the translational symmetry is broken or not, one need to investigate the bond energy oscillations in the thermodynamic limit.
Following Ref.~\cite{PhysRevB.76.195105}, we define the dimer order parameters as
\begin{eqnarray}
\Delta_D(x) &=& \langle \hat{n}_x \hat{n}_{x+1} \rangle - \langle \hat{n}_{x+1} \hat{n}_{x+2} \rangle, \nonumber \\
\Delta_G(x) &=& \sum_{\sigma} \langle \hat{c}^{\dagger}_{x,\sigma} \hat{c}_{x+1,\sigma} + h.c. \rangle - \sum_{\sigma} \langle \hat{c}^{\dagger}_{x+1, \sigma} \hat{c}_{x+2, \sigma} + h.c. \rangle, \nonumber \\
\Delta_F(x) &=& \langle {\hat{\bf S}}_x \cdot {\hat{\bf S}}_{x+1} \rangle - \langle {\hat{\bf S}}_{x+1} \cdot {\hat{\bf S}}_{x+2} \rangle,
\end{eqnarray}
where $x$ denotes the site number in one chain.
For a state with intrinsic translational symmetry breaking, these dimer order parameters will be finite in the thermodynamic limit; otherwise, they will decay and vanish.

In Figs.~\ref{BOW}(d)-\ref{BOW}(f), we plot these dimer order parameters versus site number $x$ in double-logarithmic scale, for system lengths $L_x=128, 196, 256$. 
These dimer order parameters all exhibit algebraic decay from the edge to the bulk, with the power exponents near $1/2$.
As shown in the insets, the bond order parameters at the middle of cylinder follow the linear decay to zero with $1/\sqrt{L_x}$. The same results apply for other couplings in this C0S1 state.
Therefore, we find that the bond order is not long-ranged and thus does not indicate a translational symmetry breaking. 
This charge gapped state with preserved translational symmetry is consistent with the prediction of the Lieb-Schultz-Mattis (LSM) theorem, which claims that for a ladder system with an integer number of electrons in a unit cell, it is not necessary to break the translational symmetry to open the charge gap~\cite{PhysRevLett.79.1110,PhysRevLett.78.1984,PhysRevB.58.9603}.
At the quarter filling, the electron number in each rung (unit cell) is $1$, satisfying the condition of the LSM theorem.

\begin{figure*}
   \includegraphics[width=0.9\textwidth,angle=0]{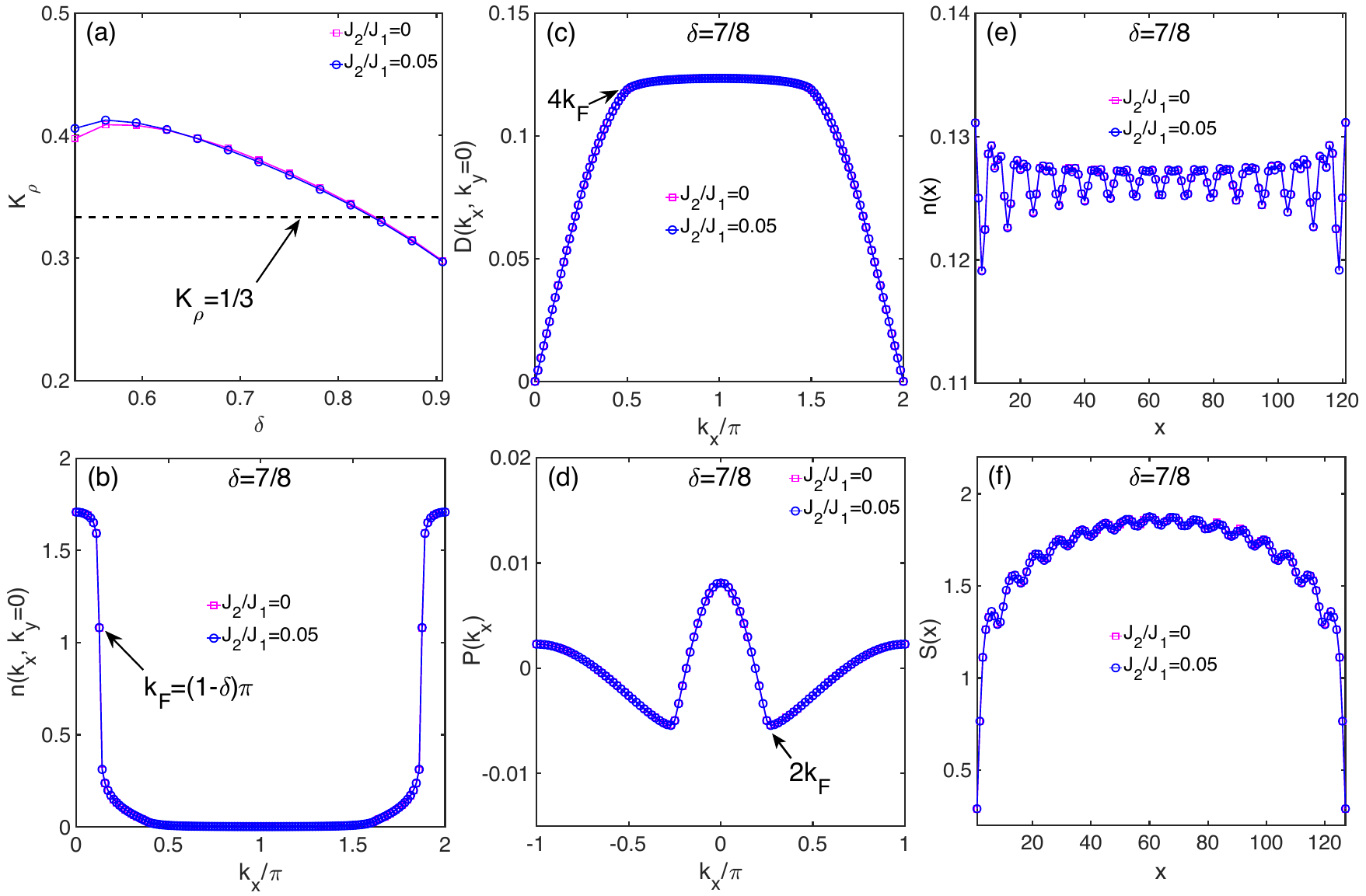}
   \caption{\label{Crossover}
   Characterization of the TLL-II regime. The results for $\delta = 7/8$, $J_2/J_1 = 0, 0.05$ are presented as examples.
   (a) Doping level dependence of the Luttinger parameter $K_{\rho}$ in the $\delta > 1/2$ region. The dotted line represents $K_{\rho}=1/3$. (b) Electron density in the momentum space with $k_y = 0$. The electrons mainly occupy a narrow momentum range near the center of the Brillouin zone. The Fermi momentum $k_F$ satisfies the relation $k_F = (1-\delta)\pi$. (c) The $k_y=0$ component of density structure factors, which show a sub-leading kink at $\mathbf{k}=(4k_F,0)$. (d) The pairing correlation structure factors with a singularity at $\mathbf{k}=(2k_F,0)$. (e) and (f) are charge density profile and entanglement entropy respectively. Both quantities have a small oscillation with the period $\lambda =1/(1-\delta)$.}
\end{figure*}

\subsection{\label{sec:over-doped}TLL-II regime}

With further increase of doping level, the system enters into a TLL-II regime that has somewhat different properties from the TLL-I regime.
The main difference is the sub-leading peak in the structure factor of density correlation function, which locates at $(2 k_F, 0)$ in the TLL-I regime but $(4 k_F, 0)$ in the TLL-II regime.
In the single-chain model, this change happens at $K_{\rho} = 1/3$~\cite{BSCS,PhysRevB.59.R2471}, which can be understood by comparing the power exponents of the $2 k_F$ mode ($1 + K_{\rho}$) and $4 k_F$ mode ($4 K_{\rho}$) in the density correlation function
\begin{equation}\label{TL-D1}
   D\left(r\right)=-\frac{K_{\rho } }{{\left(\pi r\right)}^2 }+G_1 \frac{\mathrm{cos}\left(2k_F r\right)}{r^{1+K_{\rho } } }{\mathrm{ln}}^{-\frac{3}{2}} \left(r\right)+G_2 \frac{\mathrm{cos}\left(4k_F r\right)}{r^{4K_{\rho } } },
\end{equation}
where the condition $1 + K_{\rho} = 4 K_{\rho}$ gives $K_{\rho} = 1/3$ for the shift of the sub-leading peak of density structure factor.
For the two-leg ladder, if one follows the density correlation function described by Eq.~\eqref{TL-D}, this change of sub-leading peak may happen at $K_{\rho} = 1/6$.
Following Eq.~\eqref{krho}, we first determine the Luttinger parameters $K_{\rho}$ from the derivative of density structure factor at the momentum $(0, 0)$, and the results are shown in Fig.~\ref{Crossover}(a).
$K_{\rho}$ is almost independent of the NNN couplings but decreases with doping ratio.
Furthermore, we calculate the Fermi momentum $k_F$ from the electron density in momentum space $n({\bf k})$, as shown in Fig.~\ref{Crossover}(b), which also follows the relation $k_F = ( 1 - \delta ) \pi$.
With the doping ratio dependence of $K_{\rho}$ and $k_F$, we can analyze the singularities of density structure factor.
In Fig.~\ref{Crossover}(c), we show an example of density structure factor at $\delta = 7/8$, which has the sub-leading peak at $(4 k_F, 0)$.
Therefore, by tracking the singularities of density structure factor at different doping levels, we find that the shift of the sub-leading peak still happens near $K_{\rho} = 1/3$ when $\delta \approx 83\%$ (see the dashed line in Fig.~\ref{Crossover}(a)).
This result is different from the prediction from Eq.~\eqref{TL-D}, which might be understood from the electron occupation in the momentum space.
As shown in Fig.~\ref{Crossover}(b), the electrons occupy a narrow region near the center of the Brillouin zone. 
In the perspective of tight-binding model, the few electrons in large doping will only fill one band with the lower energy, which is analog to a single chain.
For pairing correlation, we find that while $K_{\rho}$ decreases with doping ratio, the peaks of pairing structure factor in Fig.~\ref{Crossover}(d) also become weaker, showing the further suppressed pairing correlations.

We also study the charge density profile and entanglement entropy in the TLL-II regime as shown in Figs.~\ref{Crossover}(e) and \ref{Crossover}(f). We find that the period of the charge density profile and entropy is related with the doping ratio as $\lambda = 1 / (1 - \delta)$.
In addition, we find that this relationship also holds in the TLL-I regime if the doping level is commensurate (see Appendix~\ref{CDW-TLL}), which indicates that this relationship should always be satisfied in this TLL phase for commensurate doping level.
Since there are two non-uniform modes $2k_F$ and $4k_F$ in the charge sector, we may write the charge density as
\begin{equation}
n\left(x\right)=n_0 +A_{\mathrm{cdw}} \mathrm{cos}\left(2k_F x+\phi_1 \right)+B_{\mathrm{cdw}} \mathrm{cos}\left(4k_F x+\phi_2 \right).
\label{eq:cdw_TLL2} 
\end{equation}
Thus, no matter which mode is dominant, the global period of charge density profile should be compatible with the smaller momentum $2k_F$, which is nothing but $1/(1-\delta)$.

\section{\label{sec:CON} Summary and discussion}

By performing DMRG calculations, we study the quantum phase diagram of the two-leg $t$-$J$ ladder with the NNN hopping $t_2$ and spin interaction $J_2$, by tuning the doping ratio $\delta$ and $J_2/J_1$ ($t_2, J_2 > 0$, $(t_2/t_1)^2 = J_2/J_1$ and $t_1/J_1 = 3$ are fixed).
In the LEL phase at the lower doping side, the growing NNN couplings can enhance the pairing correlations, which is consistent with the observations in the wider $t$-$J$ and Hubbard models~\cite{PhysRevLett.127.097003, PhysRevLett.127.097002, PNAS.118.44,PhysRevB.106.174507,PhysRevResearch.2.033073}.
With tuning doping level and NNN couplings, the LEL phase can be distinguished as the pairing dominant ($K_{\rho} > 1/2$) and charge density dominant ($K_{\rho} < 1/2$) regimes.
At the doping level $\delta = 1/4$, the CDW state undergoes a transition to the LEL phase with closing the charge gap at very small NNN couplings.

With further increased doping level, the system enters to the TLL phase, in which the properties are almost invariant with tuning the NNN couplings.
Interestingly, we find that the TLL phase can be distinguished as two regimes, the TLL-I at the lower doping side and TLL-II at the larger doping side, which are separated at $\delta \approx 0.83$ with $K_{\rho} \simeq 1/3$.
The main difference between the two regimes is the sub-leading peak of the density correlation structure factor, which locates at the momentum $(2 k_F, 0)$ in the TLL-I regime and $(4 k_F, 0)$ in the TLL-II regime.

At $\delta = 1/2$, the system is in a C0S1 state.
With the open boundaries in this state, charge density is uniform but the horizontal bond energies have a strong dimer oscillation, which was interpreted previously as a long-range bond order wave order that breaks the translational symmetry.
We reexamine this state on larger system size and find that the dimer orders decay algebraically from the boundary to the bulk, with the power exponent $1/2$. Therefore, these dimer orders are not long-ranged and will vanish in the thermodynamic limit, which do not indicate a spontaneous translational symmetry breaking.

In the absence of the NNN couplings for the two-leg $t$-$J$ model, the LEL phase quickly emerges upon doping, and with growing doping level it changes from the pairing-dominant regime ($K_{\rho} > 1/2$) to the density-dominant regime ($K_{\rho} < 1/2$).
Compared with the DMRG results on the wider $t$-$J$ cylinders, it seems that the antiferromagnetic spin correlation increases with growing circumference at the lower doping regime and the larger doping side becomes the stripe phase on the wider system.
The density-dominant LEL regime in this two-leg system shows the enhanced density correlation with doping, which may be related to the stripe order phase on the wider systems.

Interestingly, tuning the NNN couplings $t_2/t_1>0$ can also enhance the pairing correlations in the two-leg ladder, which agrees with the observation on the wider systems. In particular, the density-dominant LEL-II can be tuned to pairing-dominant LEL-I with the growing NNN couplings, resembling the evolution from the stripe phase to the $d$-wave SC with growing $t_2/t_1$ on the wider systems.
This comparison may imply that the growing $t_2/t_1 > 0$ plays the similar role on both the two-leg and the wider systems. 
The previous study on 2-leg $t$-$J$ ladder has provided a qualitative understanding that the coherent propagation of hole pair can be enhanced by a constructive interference between $t_1$ and $t_2$, in the case of a positive $t_2/t_1$~\cite{PhysRevB.64.180513,PhysRevB.63.014414}.
Developing the microscopic theory about the enhanced pairing correlation with tuning hole dynamics on two-leg ladder should also be helpful for understanding the recent DMRG results of the $t$-$J$ model on the wider size. 

Another typical two-leg ladder is the so-called trestle lattice, which is equivalent to the 1D chain with the NNN hopping $t_2$.
For such a non-interacting system at half filling, tuning $t_2/t_1$ can change the Fermi points from two to four, which have different responses to the Hubbard interaction~\cite{Fabrizio1996,JPSJ.66.3371}. 
For a small Hubbard $U$, pairing correlation is the dominant ($K_\rho > 1/2$) in the parameter region with four Fermi points, and $K_\rho$ increases with growing $t_2/t_1 >0$ for both half-filling and doping, which can be well understood from the change of Fermi velocities in the Hartree-Fock approximation~\cite{Nishimoto2008}.
For a large $U$, although the ground state is drastically affected by interaction in the parameter region with four Fermi points, $K_\rho$ in general also increases with $t_2/t_1$ in this region~\cite{Nishimoto2008,JPSJ.76.113701}.
Notice that the studied $t$-$J$ ladder in the absence of interaction also has four Fermi points for small $t_2/t_1$ and low doping level, which may share the similar origin of the enhanced pairing correlation by tuning the NNN hopping.

\begin{acknowledgments}
We acknowledge helpful discussions with G. Roux, D. Poilblanc, and E. Orignac.
The work was supported by the National Natural Science Foundation of China 
grants (Nos. 12274014, 11834014, 11974036, 12047503, 12222412, 11874115, and 12174068), and CAS Project for Young Scientists in Basic Research (Grant No.~YSBR-057).
\end{acknowledgments}

\appendix

\renewcommand{\thefigure}{A\arabic{figure}}

\setcounter{figure}{0}

\section{\label{d-wave}Comparison of different pairing correlations}

\begin{figure*}
   \includegraphics[width=0.8\textwidth,angle=0]{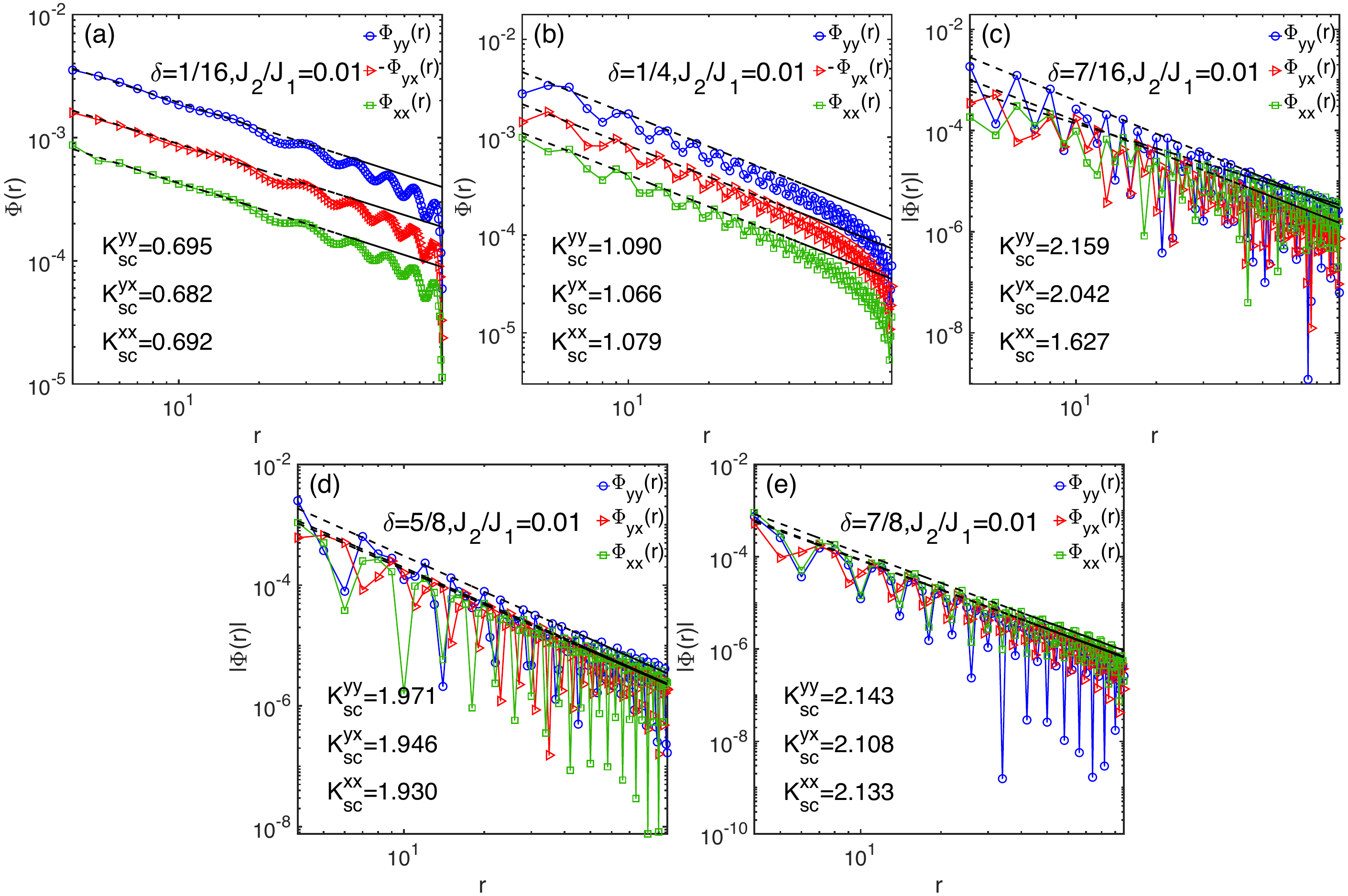}
   \caption{\label{d-wave-symmetry}Pairing correlation functions in the different regions. For simplicity, $J_{2}/J_{1}=0.01$ is fixed and five different doping ratios are chosen to represent the five regimes in the phase diagram. Various kinds of pairing correlation functions along $x$- and $y$-direction: vertical-vertical correlation $\Phi_{yy}$, horizontal-horizontal correlation $\Phi_{xx}$ and vertical-horizontal correlation $\Phi_{xy}$. The $d_{x^2-y^2}$-wave symmetry can be identified in (a) LEL-I regime and (b) LEL-II regime. (c) and (d) show the pairing correlations for the doping level before and after $0.5$ in the TLL-I regime, respectively. (e) represents the pairing correlations in the TLL-II regime.}
\end{figure*}

We have checked the pairing correlation functions between different bonds.
In Fig.~\ref{d-wave-symmetry}, we select five parameter points to represent the five regimes in the phase diagram. 
In Figs.~\ref{d-wave-symmetry}(a) and \ref{d-wave-symmetry}(b), we show that while the vertical-vertical correlations $\Phi_{yy}$ and horizontal-horizontal correlations $\Phi_{xx}$ are positive, the vertical-horizontal correlations $\Phi_{yx}$ are negative, which confirm the $d_{x^2-y^2}$-wave pairing symmetry in the LEL phase.
One can also find that although $\Phi_{yy}$ has the strongest magnitude, the different pairing correlations have the similar power exponents, namely $K_{sc}^{yy} \approx K_{sc}^{yx} \approx K_{sc}^{xx}$. 

In the TLL phase, the $d$-wave pairing symmetry is absent.
For $\delta < 0.5$ in the TLL-I regime [Fig.~\ref{d-wave-symmetry}(c)], the power exponents of the different pairing correlations have slight differences.
In the other regimes [Figs.~\ref{d-wave-symmetry}(d) and \ref{d-wave-symmetry}(e)], they are still consistent.
Therefore, in the main text we demonstrate the results of $\Phi_{yy}$ as the representative.

\renewcommand{\thefigure}{B\arabic{figure}}

\setcounter{figure}{0}

\section{\label{Stru-fac-LEL}Structure factors in the LEL phase}

\begin{figure*}
   \includegraphics[width=1.0\textwidth,angle=0]{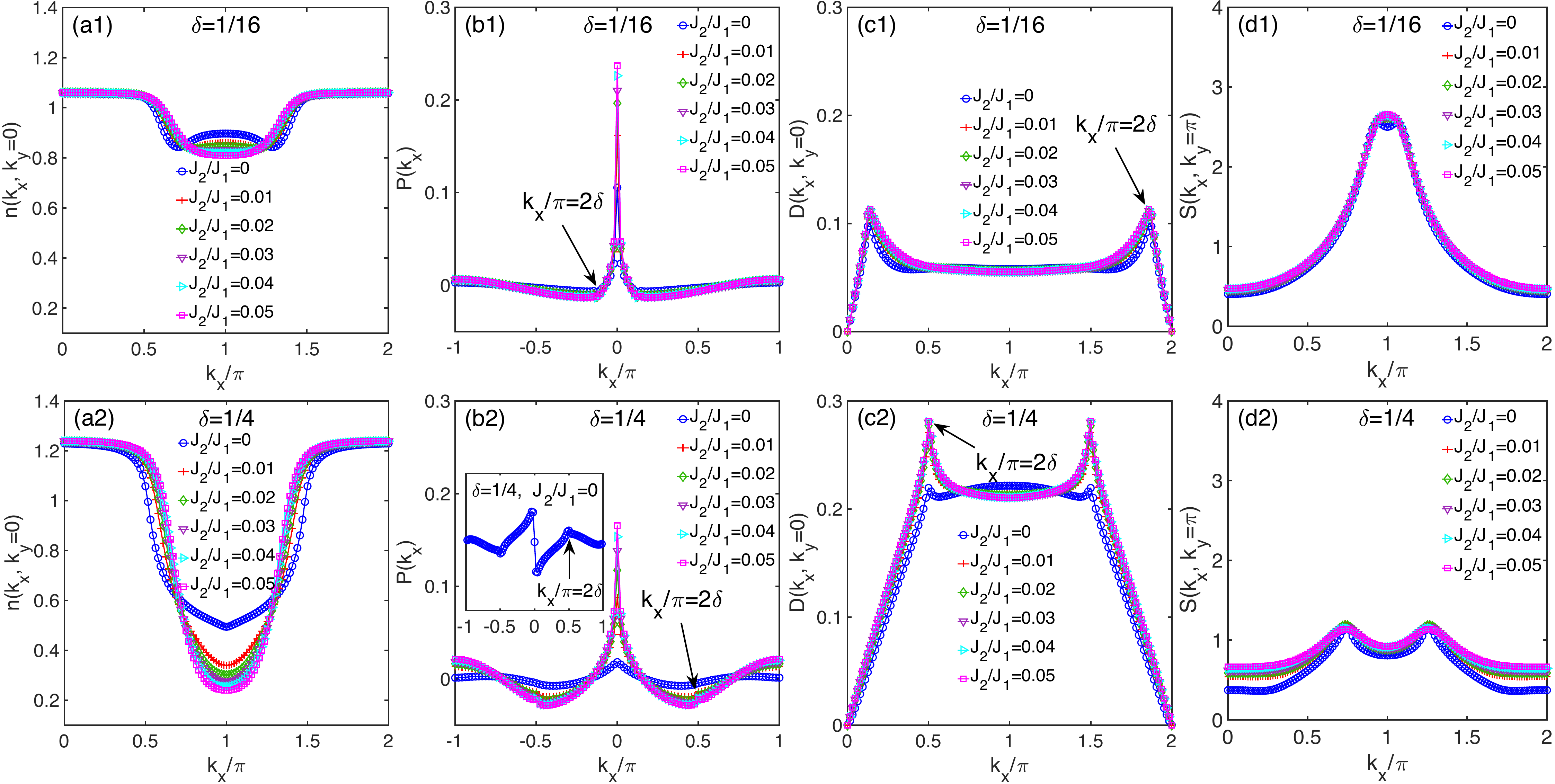}
   \caption{\label{Structure-factors-LEL}Structure factors in the LEL phase.
   (a1)-(d1) are the single-electron occupation probability, pairing structure factor, 
   density structure factor, and spin structure factor with different $J_{2}/J_{1}$ for $\delta=1/16$. (a2)-(d2) represent the same types structure factors for $\delta=1/4$. Some symbolic peaks and kinks are marked in the figures, which are determined by the doping level.
   The inset in (b2) is the derivative of $P(k_x)$ for $J_{2}/J_{1}=0$ at $\delta=1/4$.}
\end{figure*}

The structure factors in the LEL phase are shown in Fig.~\ref{Structure-factors-LEL}. 
The electron densities in the momentum space $n({\bf k})$ are shown in Figs.~\ref{Structure-factors-LEL}(a1) and \ref{Structure-factors-LEL}(a2). 
The structure factors of pairing correlation $P(k_x)$ in Figs.~\ref{Structure-factors-LEL}(b1) and \ref{Structure-factors-LEL}(b2), using the Fourier transformation of $\Phi_{yy}$, show a strong singular peak at $k_x = 0$, which is consistent with enhanced pairing correlation. Besides, $P(k_x)$ also has a small kink at $k_x = 2\delta \pi$ that reflects the period of pairing correlation $\lambda=1/\delta$.
To show the kink clearly, we demonstrate the derivative of $P(k_x)$ for $J_{2}/J_{1}=0$, $\delta=1/4$ in the inset of Fig.~\ref{Structure-factors-LEL}(b2).
The structure factor of density correlation $D({\bf k})$ in the LEL phase also shows singular peak at $\mathbf{k}=(2\delta \pi, 0)$ [Figs.~\ref{Structure-factors-LEL}(c1) and \ref{Structure-factors-LEL}(c2)].
The peak value increases with growing doping level, indicating that doping will enhance density correlation.  
In Figs.~\ref{Structure-factors-LEL}(d1) and \ref{Structure-factors-LEL}(d2), we show the spin structure factor. 
For small $J_2/J_1$ at half filling, spin structure factor has a peak at ${\bf k} = (\pi, \pi)$. 
With doping, the peak splits and move away from ${\bf k} = (\pi, \pi)$.

\renewcommand{\thefigure}{C\arabic{figure}}

\setcounter{figure}{0}

\section{\label{Kc-CDW} Extracting the exponent $K_{\mathrm{c}}$ by fitting charge density profile}

\begin{figure*}
   \includegraphics[width=0.9\textwidth,angle=0]{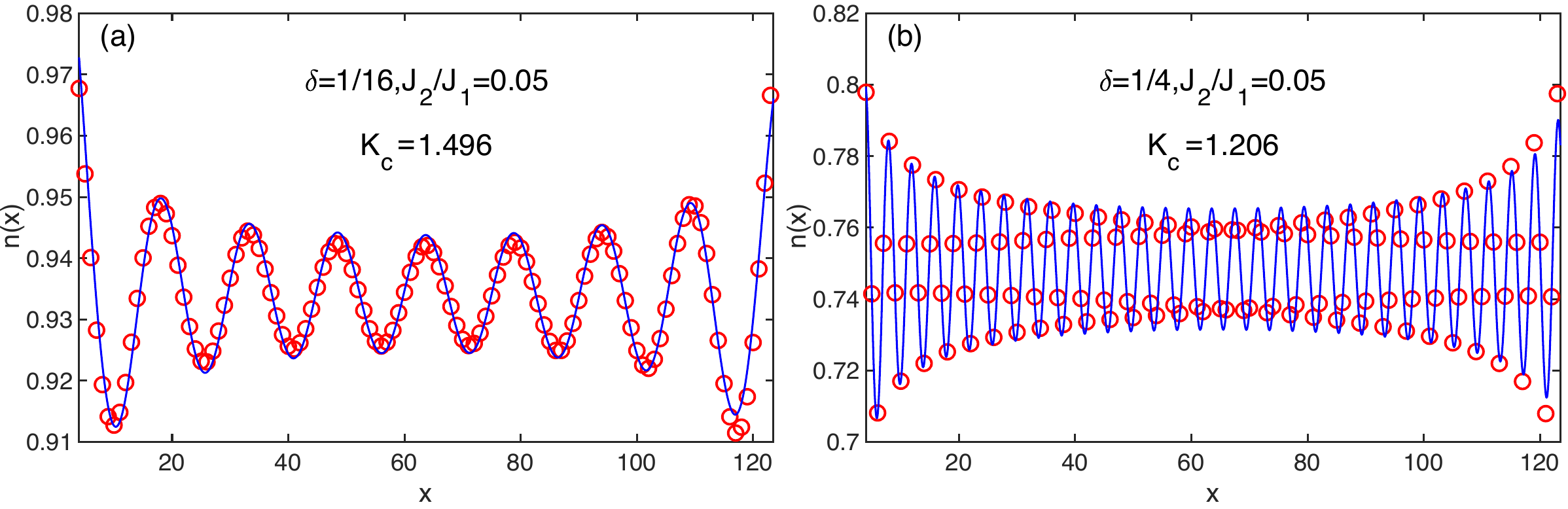}
   \caption{\label{CDW-Kc}Charge density profiles in the LEL phase. The charge density distributions $n_x =\frac{1}{L_y }\sum_{y=1}^{L_y } \left\langle {\hat{n} }_{x,y} \right\rangle$ with $J_{2}/J_{1}=0.05$ for (a) $\delta=1/16$, and (b) $\delta=1/4$. The blue lines are the fitting curves to the function $n\left(x\right)=n_0 +A_{\mathrm{cdw}} \mathrm{cos}\left(2k_F x+\phi \right)$, where $A_{\mathrm{cdw}}=A_{0}[x^{-K_c /2} +{\left(L_x +1-x\right)}^{-K_c /2}]$ and $k_F$ are the amplitude and Fermi momentum, $\phi$ is a phase shift.}
\end{figure*}

In the open boundary system, due to the Friedel oscillation of charge density, the power exponent $K_{\mathrm{c}}$ can also be obtained by fitting the charge density profile using the formula,
\begin{equation}
n\left(x\right)=n_0 +A_{\mathrm{cdw}} \mathrm{cos}\left(2k_F x+\phi \right),
\end{equation}
where $A_{\mathrm{cdw}}=A_{0}[x^{-K_c /2} +{\left(L_x +1-x\right)}^{-K_c /2}]$, $k_F$ and $\phi$ are fitting parameters.
Here, we choose the systems with $J_{2}/J_{1}=0.05$ for demonstration [Fig.~\ref{CDW-Kc}]. We eliminate some data points at the edges to avoid boundary effect. The fitted $K_{\mathrm{c}}$ in Fig.~\ref{CDW-Kc} agree with the results we obtained by fitting the decay of density correlation function.

\renewcommand{\thefigure}{D\arabic{figure}}

\setcounter{figure}{0}

\section{\label{central-charge and entropy}The evolution of entanglement entropy and central charge at $1/4$ doping}

\begin{figure*}
   \includegraphics[width=0.65\textwidth,angle=0]{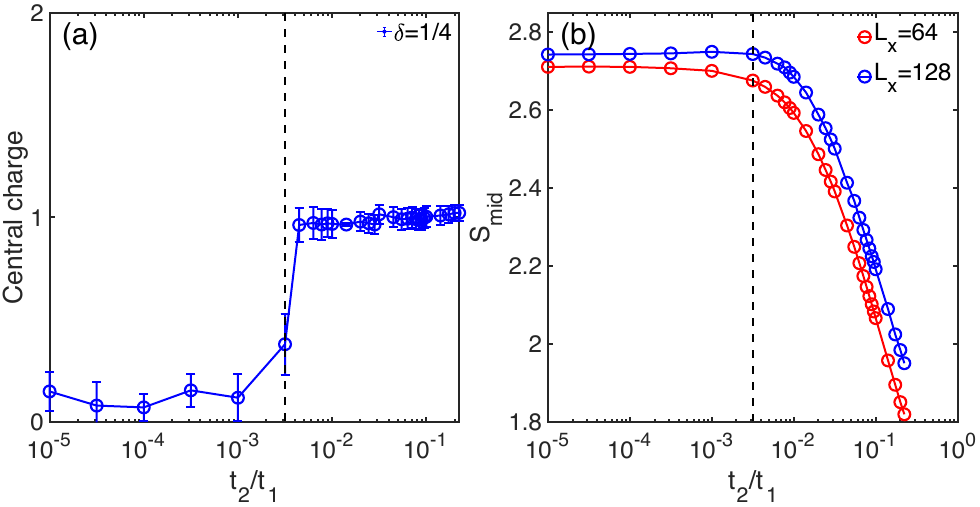}
   \caption{\label{KT-transition}
   Phase transition from the CDW to the LEL state at $1/4$ doping. (a) $t_2/t_1$ dependence of the obtained central charge on the $L_x = 128$ ladder, which changes from $c \approx 0$ to $c \approx 1$ near $t_{2}/t_{1} \sim 0.003$. Notice here the variable $t_2/t_1$ is shown as the logarithmic scale. (b) $t_2/t_1$ dependence of the entanglement entropy measured in the middle of the ladders with $L_x = 64$ and $128$, where the turning point is still around $t_{2}/t_{1} \sim 0.003$.}
\end{figure*}

We fit the central charge at $\delta = 1/4$ with growing $t_2/t_1$.
As shown in Fig.~\ref{KT-transition}(a), the fitted central charge exhibits a change from $c \approx 0$ to $c \approx 1$ near $t_{2}/t_{1} \sim 0.003$. 
Notice that the $x$-axis variant $t_2/t_1$ is labeled by the logarithmic scale. 
One can also find that the entanglement entropy at the middle of ladder $S_{\mathrm{mid}}$ in Fig.~\ref{KT-transition}(b) changes slightly with $t_{2}/t_{1}$ in the CDW phase but decreases gradually in the LEL phase, consistent with the transition observed from the change of central charge.
This transition happens at very small NNN couplings, which is consistent with the tiny charge gap at $t_2/t_1 = 0$~\cite{PhysRevB.65.165122,PhysRevB.76.195105}.

\renewcommand{\thefigure}{E\arabic{figure}}

\setcounter{figure}{0}

\section{\label{correlation-TLL}Correlation functions in the TLL-I regime}

\begin{figure*}
   \includegraphics[width=0.85\textwidth,angle=0]{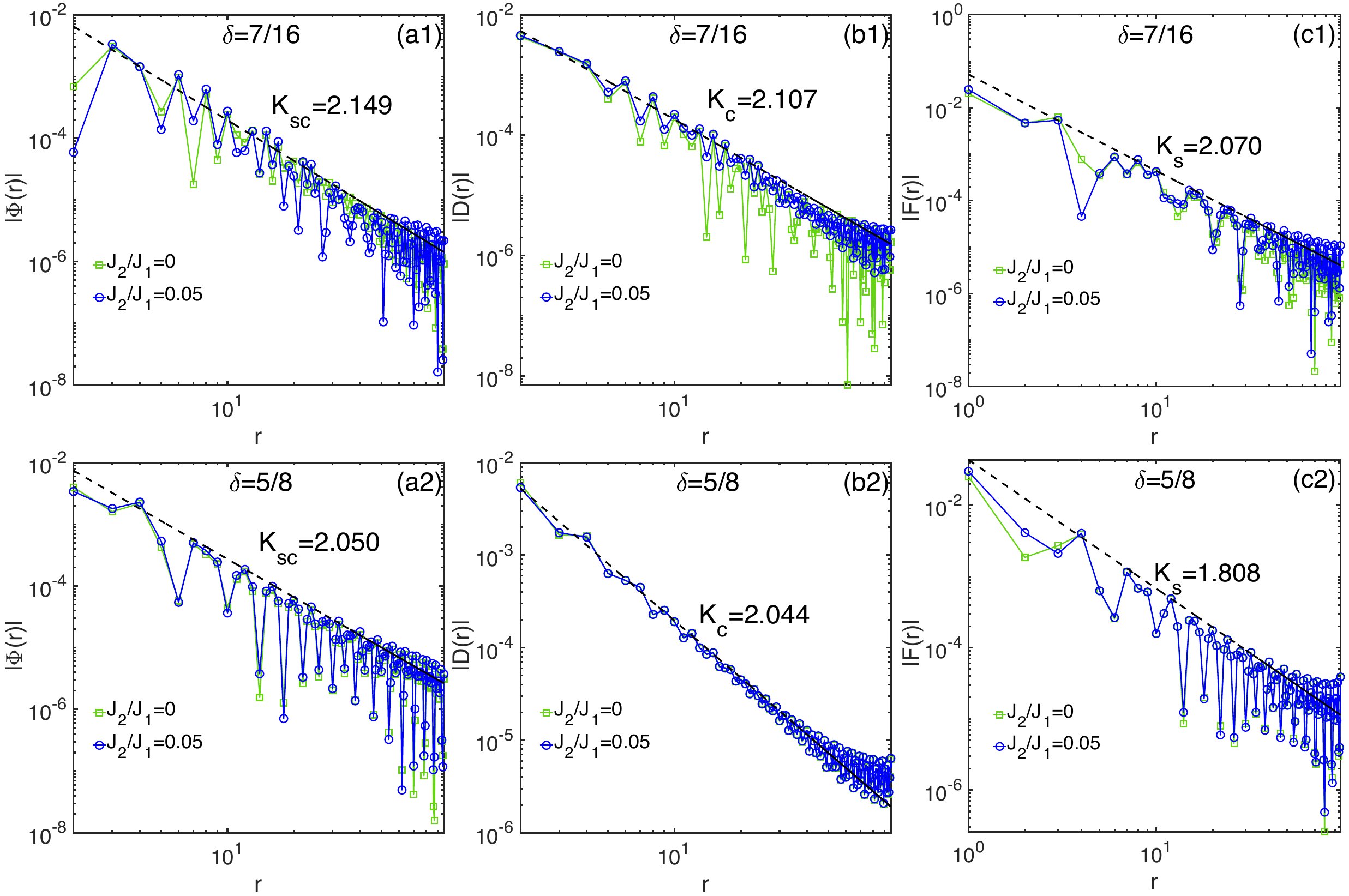}
   \caption{\label{TLL-regime}Correlation functions in the TLL-I regime.
   The doping ratios $\delta=7/16$ and $\delta=5/8$ are chosen as representatives to represent the TLL-I regime before and after $\delta=0.5$. (a1)-(c1) are pairing correlation, density correlation and spin correlation function for $\delta=7/16$ with different $J_{2}/J_{1}$ on double-logarithmic
   sacle, $K_{\mathrm{sc}}$. $K_{\mathrm{c}}$ and $K_{\mathrm{s}}$ are fitted for $J_{2}/J_{1}=0.05$. (a2)-(c2) are the similar results for $\delta=5/8$.}
\end{figure*}

In this section we choose $J_{2}/J_{1} = 0.05$ as the representative to show the power exponents of correlation functions in the TLL-I regime with $\delta = 7/16$ for $\delta < 0.5$ and $\delta = 5/8$ for $\delta > 0.5$ [Fig.~\ref{TLL-regime}].

Following Eq.~\eqref{krho}, we first determine the Luttinger parameters for $\delta=7/16$ and $\delta=5/8$ as $K_{\rho}\approx0.440$ and $K_{\rho}\approx0.404$, respectively.
For the pairing correlation functions in Figs.~\ref{TLL-regime}(a1) and \ref{TLL-regime}(a2), the power exponents are fitted as $K_{\mathrm{sc}} \approx 2.149$ and $K_{\mathrm{sc}} \approx 2.050$. For $\delta=5/8$, the power exponent $K_{\mathrm{sc}}$ is consistent with the second term in Eq.~(\ref{TL-Phi}), i.e. $2K_{\rho }+1/(2K_{\rho })$. 
For density correlation functions [Figs.~\ref{TLL-regime}(b1) and \ref{TLL-regime}(b2)], the power exponents are all about $2$, indicating that the first term in Eq.~(\ref{TL-D}) is the main contribution. 
For spin correlation, we obtain the power exponents $K_{\mathrm{s}}\approx2.070$ and $K_{\mathrm{s}}\approx1.808$ [Figs.~\ref{TLL-regime}(c1) and \ref{TLL-regime}(c2)], which are consistent with the first term ($2$) and the second term ($1+2K_{\rho }$) of Eq.~(\ref{TL-F}), respectively.  

\renewcommand{\thefigure}{F\arabic{figure}}

\setcounter{figure}{0}

\section{\label{CDW-TLL}Charge density profile in the TLL phase}

\begin{figure*}
   \includegraphics[width=0.85\textwidth,angle=0]{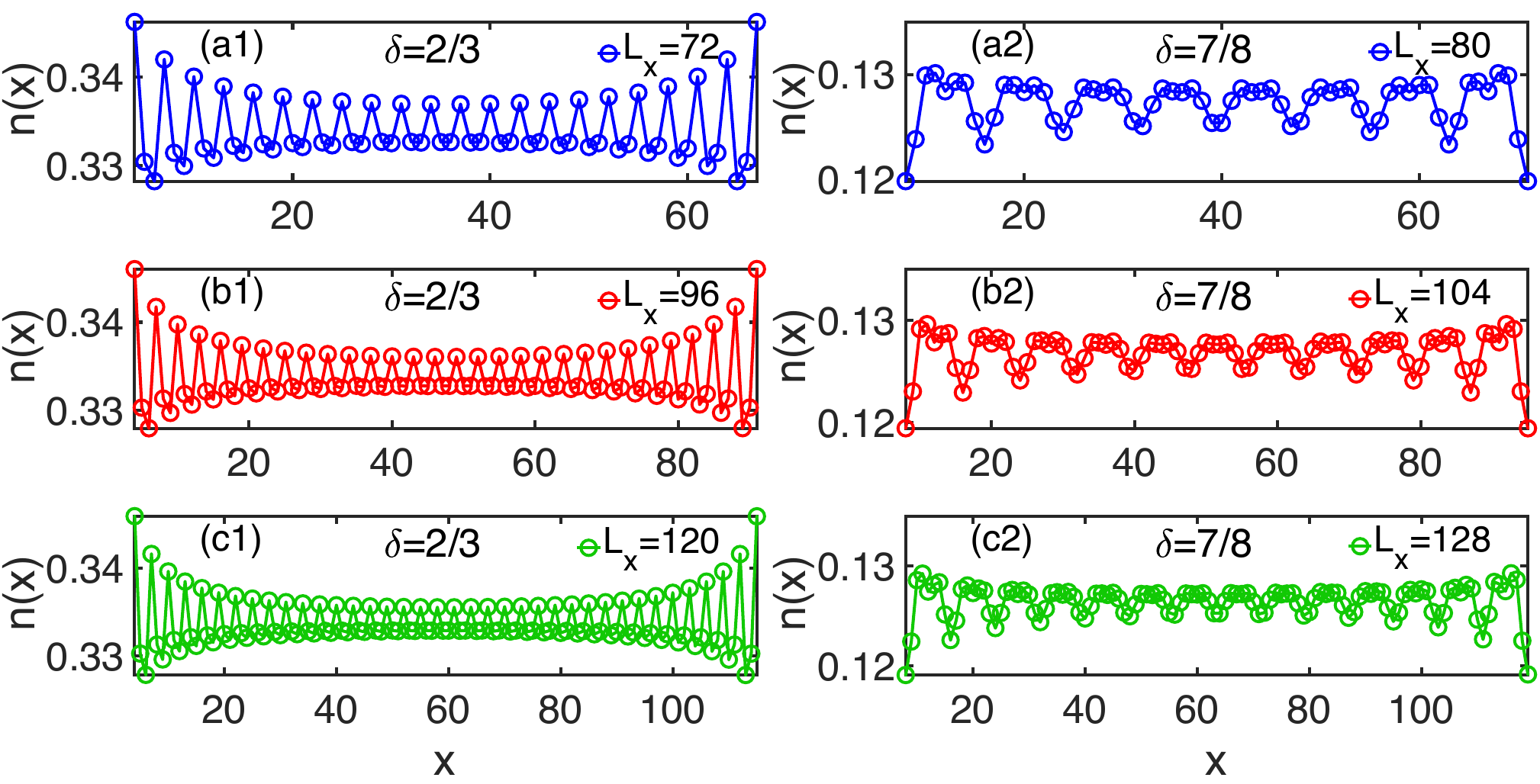}
   \caption{\label{TLL-CDW}Charge density profile in the TLL phase.
   Two doping ratios $\delta=2/3$ and $\delta=7/8$ are chosen as representatives to characterize the TLL-I and TLL-II regime. (a1)-(c1) demonstrate the results for $\delta=2/3$ with different lattice sizes, all of which have a period of $\lambda=3$. (a2)-(c2) are the similar figures for $\delta=7/8$ with a period of $\lambda=8$.}
\end{figure*}

We show the charge density profile for the TLL phase in Fig.~\ref{TLL-CDW}. Two doping ratios $\delta=2/3$ and $\delta=7/8$ are selected to represent the TLL-I and TLL-II regime. In Figs.~\ref{TLL-CDW}(a1)-\ref{TLL-CDW}(c1), we show the charge density distributions at different lattice sizes which all match the doping level $\delta=2/3$. The DMRG results show that the period of the CDW is $\lambda=3$, which strictly obeys the relation $\lambda = 1 / (1 - \delta)$. Similar conclusion also can be drawn in the TLL-II regime as shown in Figs.~\ref{TLL-CDW}(a2)-\ref{TLL-CDW}(c2) with the period $\lambda=8$. 
Notice that if the system length is not compatible with doping level, charge density profile does not show a clear period. 

\clearpage
\bibliography{refs}
\end{document}